\documentclass{emulateapj}
\usepackage{subfigure}
\usepackage{graphicx}
\shorttitle{Self-Regulated Feedback: Thermal Instabilities and Multiphase gas}
\shortauthors{Gaspari, Ruszkowski \& Sharma}

\def\gsim{\;\rlap{\lower 2.5pt
 \hbox{$\sim$}}\raise 1.5pt\hbox{$>$}\;}
\def\lsim{\;\rlap{\lower 2.5pt
   \hbox{$\sim$}}\raise 1.5pt\hbox{$<$}\;}

\newcommand\msun{\rm{M_{\odot}}}

\def\stacksymbols #1#2#3#4{\def\theguybelow{#2}
        \def\verticalposition{\lower#3pt}
        \def\spacingwithinsymbol{\baselineskip0pt\lineskip#4pt}
        \mathrel{\mathpalette\intermediary#1}}
\def\intermediary #1#2{\verticalposition\vbox{\spacingwithinsymbol
        \everycr={}\tabskip0pt
        \halign{$\mathsurround0pt#1\hfil##\hfil$\crcr#2\crcr
                \theguybelow\crcr}}}

\def\gta{\stacksymbols{>}{\sim}{2.5}{.2}}

\begin{document}

\title{Cause and Effect of Feedback: Multiphase Gas in Cluster Cores Heated by AGN Jets}
\author{M. Gaspari$^{1,6}$, M. Ruszkowski$^{2,3}$ and P. Sharma$^{4,5}$}
\affil{$^{1}$Department of Astronomy, University of Bologna, Via Ranzani 1, Bologna 40127, Italy; e-mail: massimo.gaspari4@unibo.it (MG)\\
$^{2}$Department of Astronomy, University of Michigan, 500 Church Street, Ann Arbor, MI 48109, USA; e-mail: mateuszr@umich.edu (MR)\\
$^{3}$The Michigan Center for Theoretical Physics, 3444 Randall Lab, 450 Church St, Ann Arbor, MI 48109, USA\\
$^{4}$Astronomy Department and Theoretical Astrophysics Center, 601 Campbell Hall, University of California, Berkeley, CA 94720, USA; \\
$^{5}$Department of Physics, Indian Institute of Science, Bangalore, India 560012; e-mail: prateek@physics.iisc.ernet.in (PS)}
\altaffiltext{6}{Work performed while on leave at the Depertment of Astronomy in the University of Michigan, Ann Arbor.}

\begin{abstract}
\noindent
Multiwavelength data indicate that the X-ray emitting plasma in the cores of galaxy clusters is not cooling catastrophically. To large extent, cooling is offset by heating due to active galactic nuclei (AGN) via jets. The cool-core clusters, with cooler/denser plasmas, show multiphase gas and signs of some cooling in their cores. These observations suggest that the cool core is locally thermally unstable while maintaining global thermal equilibrium. Using high-resolution, three-dimensional simulations we study the formation of multiphase gas in cluster cores heated by highly-collimated bipolar AGN jets. Our key conclusion is that spatially extended multiphase filaments form only when the instantaneous ratio of the thermal instability and free-fall timescales ($t_{\rm TI}/t_{\rm ff}$) falls below a critical threshold of $\approx 10$. When this happens, dense cold gas decouples from the hot ICM phase and generates inhomogeneous and spatially extended H$\alpha$ filaments. These cold gas clumps and filaments `rain' down onto the central regions of the core, forming a cold rotating torus and in part feeding the supermassive black hole. Consequently, the self-regulated feedback enhances AGN heating and the core returns to a higher entropy level with $t_{\rm TI}/t_{\rm ff} > 10$. Eventually the core reaches quasi-stable global thermal equilibrium, and cold filaments condense out of the hot ICM whenever $t_{\rm TI}/t_{\rm ff} \lesssim 10$. This occurs despite the fact that the energy from AGN jets is supplied to the core in a highly anisotropic fashion. The effective spatial redistribution of heat is enabled in part by the turbulent motions in the wake of freely-falling cold filaments.  Increased AGN activity can locally reverse the cold gas flow, launching cold filamentary gas away from the cluster center. Our criterion for the condensation of spatially extended cold gas is in agreement with observations and previous idealized simulations.
\end{abstract}

\keywords{cooling flows -- galaxies: active -- galaxies: clusters: general -- galaxies: jets -- intergalactic medium --  methods: numerical}

\section{Introduction}

Unopposed radiative cooling of the intracluster medium (ICM) would lead to the cooling catastrophe -- 
an inflow of gas toward the center, due to the loss of pressure, at the rate of 
several hundred $\msun$ yr$^{-1}$ (see \citealt{fab94} for a review). 
The main consequences would be strongly peaked surface brightness maps, 
central temperatures falling below $T_{\rm vir}/3$, and a huge accumulation of cold gas in the
cluster core, well over $10^{12}$ $\msun$.
The fact that none of these cooling flow features are observed  
(e.g., \citealt{pet01,pet03,tam03,pef06}) points out that either the cooling proceeds non-radiatively or it is balanced by some kind of heating. 

The most popular and promising heating mechanism is the active galactic nuclei (AGN) feedback. Radio and X-ray data provide overwhelming evidence for the interaction of jets/outflows from the supermassive black holes with the surrounding ICM (\citealt{boe93,bla01,fij01,jon02,raf06,mcn07,fin08,raf08}). The power supplied by the AGN is in the range of $10^{43}-10^{46}$ erg s$^{-1}$, 
adequate to globally offset radiative cooling losses.
Further support for the outflow model comes from blueshifted absorption lines in the core of several galaxies: in the optical band (\citealt{nes08,nes11}), UV and X-ray bands (\citealt{tom10}; \citealt{cre03} for a review), and at 21-cm (\citealt{mor05,mor07}). These lines originate at kpc distances and correspond to velocities of up to several $10^4$ km s$^{-1}$ and mass outflow
rates around a $\msun$ yr$^{-1}$ or more. The geometry is still unclear, but radio images are suggestive of conical outflows with narrow opening angles and possibly consistent with precession.

While the global energy balance may be achieved with the help of the AGN feedback and/or other mechanisms \citep{rub02, brm03, gor08, coo08, ruo11}, the issues regarding the coupling of 
the mechanical AGN energy to the thermal ICM energy, along with the spatial distribution of heating, are not settled.
The coupling may occur, for example, via turbulent mixing of AGN bubbles with the ICM \citep{scb08}, 
viscous dissipation of weak shocks \citep{rusz04a,rusz04b}, and cosmic ray heating \citep{guo08, rusz08,sha09}. 
Other mechanisms may also help heat the ICM and redistribute the energy throughout the cool core:
dynamical friction \citep{ezk04}, thermal conduction \citep{nm01,bog09,ruo10,ruo11,par10}, or sloshing motions \citep{mor10,zuhone10}. 

For most microscopic heating mechanisms 
the ICM is expected to be locally thermally unstable.
In other words, even if the ICM maintains global thermal balance without a cooling flow, some of the gas can experience runaway cooling.
Indeed, the observations by \citet{edg01} and \citet{sal03} revealed that cooling flows
are not totally stopped: the mass of cold gas, mainly dominated by molecular H$_2$ 
associated with CO emission, reaches $10^{9}-3\times10^{11}$ $\msun$ in
the majority of cool-core clusters. Cooling may also be inferred from star formation
(e.g., O'Dea 2008, typically $1-10$ $\msun$ yr$^{-1}$).
More recently, a wide survey conducted by \citet{mcd10,mcd11a,mcd11b}
utilized H$\alpha$ emission as a robust tracer of the cold phase. In more than
60\% of clusters and groups, 
extended and/or nuclear cold gas was detected. The morphology of this phase
is usually clumpy or filamentary and elongated toward the center (e.g., in the Perseus cluster; \citealt{sal06,fab08}).

\citet{cav08}, \citet{voi08} and \citet{mcd11a} have shown the existence of a tight correlation between the
amount of cold gas and the ICM entropy or the cooling time. 
Only clusters with central entropies below 30 keV cm$^2$ show evidence for the extended cold gas. This criterion can be expressed in terms of the ratio of the cooling time ($t_{\rm cool}$\footnote{The cooling time and the thermal instability timescale are similar for a microscopic heating mechanism which is independent of density, so we use $t_{\rm TI}$ and $t_{\rm cool}$ interchangeably.}) and the free-fall time ($t_{\rm ff}$).

Analytical estimates and 
Cartesian simulations by \citeauthor{mcc11} (\citeyear{mcc11} -- M11) pointed out that a
spatially extended and inhomogeneous cold phase can form whenever $t_{\rm cool}/t_{\rm ff}\la 1$. 
Previous studies (e.g., \citealt{bas89}) did not predict the presence of multiphase gas because they neglected a spatially distributed heating, which provides global thermal stability.
Fig.~12 in M11, based on observational data, 
suggests that the critical threshold is actually higher, $\approx10$. 
\citeauthor{sha11} (\citeyear{sha11} -- S11) extended the work of M11, considering 
spherical feedback models. They found that, due to the
geometrical compression, the criterion is in fact less stringent:
instability occurs when $t_{\rm cool}/t_{\rm ff}\la 10$.

The results presented in M11 and S11 are very promising, but
the distributed heating models adopted there are idealized. In these models 
the heating is either added using a prescribed spherically-symmetric 
functional form, or it is constructed to be equal to the shell-averaged radiative cooling loss at every radius.
Given the simplicity of these prescriptions, it is necessary to consider more realistic heating models to reliably study the formation of the multiphase medium, along with reproducing several important features, such as shocks, cavities and turbulence.
We perform high-resolution 
three-dimensional simulations which incorporate ICM heating via collimated AGN jets. 
This study is also a natural extension of our previous efforts (\citealt{gas09,gas11a,gas11b} -- G11a,b), which included AGN jet feedback but mainly focused on the dynamics of the hot ICM phase, suppressing the formation of the multiphase gas via cold phase dropout.

Our more realistic AGN feedback simulations confirm that  extended multiphase ICM forms only when the {\it instantaneous} ratio of the thermal instability and free-fall timescales falls below a critical threshold of $\approx 10$, irrespective of the initial condition. The cluster core attains rough thermal equilibrium after a few cooling times. In addition to occasional extended cold filaments, centrally concentrated cold gas exists at most times. Thus AGN feedback primarily occurs via the cold phase.
Our simulations also show that sometimes the cold gas can be carried out by AGN jets. The evidence for such cold filaments dredged out from the central galaxy is seen in  the Perseus cluster \citep{fab03}. 
\\

\section[]{Numerical methods}

We use the adaptive mesh refinement code FLASH 4.0 (\citealt{fry00}),
modified to simulate mechanical AGN feedback in the cooling ICM. 
The numerical finite-volume method employed is the third-order accurate
split PPM, suited to solve the Euler equations including the required heating and cooling terms:

\begin{equation}\label{cont}
\frac{\partial\rho}{\partial t} + {\bf\nabla}\cdot\left(\rho {\bf v}\right) = \pm{\bf\nabla}\cdot{(\dot{m}_{\rm jet}\,{\bf n}_{\rm z})},
\end{equation}
\begin{equation}\label{mom}
\frac{\partial\rho {\bf v}}{\partial t} + {\bf \nabla}\cdot\left(\rho {\bf v} \otimes {\bf v}\right) + {\bf \nabla}{P} = \rho{\bf g}\, \pm{\bf\nabla}\cdot({\dot{m}_{\rm jet} \bf{n}_{\rm z}\otimes{\bf v}_{\rm jet}}),
\end{equation}
\[
\frac{\partial\rho \varepsilon}{\partial t} + {\bf\nabla}\cdot\left[\left(\rho \varepsilon + P\right) {\bf v}\right] = \rho{\bf v}\cdot{\bf g} -\: n_{\rm{e}} n_{\rm{i}} \Lambda(T,Z)
\]
\begin{equation}\label{ene}
\;\quad\quad\quad\quad\quad\quad\quad\quad\quad\quad\quad\quad  \pm{\bf\nabla}\cdot{\left(
\frac{1}{2}\dot{m}_{\rm jet} v_{\rm jet}^2 \;{\bf n}_{\rm z}\right)},
\end{equation}
\begin{equation}\label{eos}
P = \left(\gamma -1\right)\rho \left(\varepsilon-\frac{1}{2}v^2\right),
\end{equation}

\noindent
where $\rho$ is the gas density, $\bf{v}$ the velocity, $\varepsilon$ the specific 
total energy (internal plus kinetic), $P$ the pressure, and 
$\gamma = 5/3$ the adiabatic index. 
The temperature is computed from $P$ and $\rho$ using Eq.~(\ref{eos}), 
with an atomic weight $\mu\simeq 0.62$, appropriate for a totally
ionized plasma with 25\% He in mass.
The externally imposed gravitational
acceleration ${\bf g}$ accounts for the dark matter NFW halo
\citep{nfw96} characterized by a virial mass and radius of  $10^{15}$ M$_\odot$
and 2.6 Mpc, respectively (concentration parameter $\simeq 6.6$). 
Our implementations of the cooling and jet flux terms in Equations (\ref{cont}) -- (\ref{ene}) 
deserve careful discussion and are presented in Sections 2.1 and 2.2, respectively.

We decided to neglect magnetic fields and anisotropic thermal conduction.
M11 showed that the anisotropic conduction changes the morphology of the cold gas
(more filamentary), 
but does not affect the criterion for the generation of the multiphase gas due to thermal instabilities.
We defer the study of these effects to future work.\\

\subsection[]{Radiative Cooling}

The intracluster medium emits radiation mainly in the X-ray band
due to Bremsstrahlung ($T\gta10^7$ K) and due to
line emission at lower temperatures ($10^4$ -- $10^7$ K). Assuming collisional ionization
equilibrium, the cooling source term in Eq.~(\ref{ene}) can be modeled
with a cooling function $\Lambda(T,Z)$: we use an analytical
fit to the \citet{sud93} tabulated values for a fully ionized plasma with metallicity
$Z = 0.3$ Z$_\odot$ (e.g., \citealt{tam04}), setting the temperature floor at $10^4$ K. 
The total emissivity of the gas is also proportional
to the electron and ion number densities: $\mathcal{L}=n_{\rm{e}} n_{\rm{i}} \Lambda(T,Z)$. For simplicity
we assume that molecular weights are constant, with $\mu_e\mu_i=1.464$.

An important difference between the current implementation of the cooling term and that adopted in our previous models (G11a,b; \citealt{brm02}) is the absence of the dropout term, which removes the cold gas from the domain. Here we are interested in the formation and growth of thermal instabilities, producing cold clumps and filaments in the ICM.

The presence of the cooling term can significantly slow down the computation. 
Explicit schemes, such as Runge-Kutta, are usually simple and fast (per timestep),
but impose severe restrictions on the timestep $\Delta t$, making it much smaller than the 
regular hydrodynamical timestep. Implicit schemes allow to use larger timesteps, but these methods
are less precise.
In a splitting framework, i.e., first solving the hydrodynamic 
PDE and then the cooling source term ODE (assuming fixed density),
no implicit or explicit solver is actually needed for the ODE step.

The new temperature $T_j^{n+1}$ of a cell $j$ at time ${n+1}$ can
be computed from the old temperature $T_j^{n}$
by exactly evaluating the following integral:
\begin{equation}\label{exact}
\int^{T_j^{n+1}}_{T_j^n}{\frac{dT'}{\Lambda(T')}}=-\frac{(\gamma-1)\mu}{k_{\rm B}\mu_{\rm e}\mu_{\rm i}m_{p}}\rho_j\Delta t\:.
\end{equation}
\noindent
A practical way to compute the integral is to split the integration interval into two parts:
one from $T_{j}^n$ to $T_{\rm ref}$ and the other one from $T_{\rm ref}$ to $T_{j}^{n+1}$,
using an arbitrary reference value (see \citealt{tow09}). This leads to \\
\noindent
\begin{equation}\label{exact2}
T^{n+1}_j=Y^{-1}\left[Y(T^n_j)+\frac{T^n_j}{T_{\rm ref}}\frac{\Lambda(T_{\rm ref})}{\Lambda(T^n_j)}\frac{\Delta t}{t_{\rm cool}}  \right],
\end{equation}

\noindent
where the cooling time and $Y$ are defined as:
\begin{equation}\label{tcool}
t_{\rm cool}\equiv\frac{e}{n_{e}n_{i}\Lambda(T_{j}^{n})}=\frac{k_{\rm B} \mu_{\rm e} \mu_{\rm i}m_{\rm p}}{(\gamma-1)\mu}\frac{T^n_j}{\rho_j\Lambda(T_j^n)},
\end{equation}
\begin{equation}\label{Y}
Y(T)\equiv\frac{\Lambda(T_{\rm ref})}{T_{\rm ref}}\int^{T_{\rm ref}}_T\frac{dT'}{\Lambda(T')},
\end{equation}
where $e \equiv P/(\gamma-1)$ is the internal energy density.
The temperature $T_{j}^{n+1}$ can be exactly computed 
if the reciprocal of the cooling function is analytically integrable.
Thus, we decompose the cooling function into a large number of piece-wise powerlaws in logarithmic temperature intervals
\begin{equation}
\Lambda(T)=\Lambda_k\left(\frac{T}{T_k}\right)^{\alpha_k}\quad\quad\quad\quad\quad T_k\leq T\leq T_{k+1}
\end{equation}

\noindent
and precompute at initialization the constants of integration for each interval. The integral in Eq.~(\ref{Y}) can be computed analytically using this
powerlaw decomposition of the cooling function. The only effort in retrieving $T_{j}^{n+1}$ comes from finding the correct interval
in which the argument of $Y^{-1}$ resides (Eq.~(\ref{exact2})).

Aside from being exact (in the ODE step), this method is faster than either explicit or implicit methods.
The error from the piece-wise powerlaw approximation is negligible, because
the number of the temperature intervals can be made very large (matching the tabulated Sutherland \& Dopita values), without large memory requirements.

In principle, our solver does not require a timestep limiter for stability. However,
the coupling between the hydrodynamics and source term integration is still approximate.  Thus we preferred 
to impose an upper limit on the cooling timestep ($\sim t_{\rm cool}$).
We note that this did not dramatically increase the computational requirements.
In order to further improve the coupling, we have also implemented a second order accurate
temporal evolution in the splitting scheme, i.e., 
we alternate the order of the integration of the hydrodynamical (H) and source terms (S) as follows: H-S-S-H. 

We have compared the explicit calculation, setting a strong limiter ($\la0.1$ $t_{\rm cool}$), 
with the `exact'
cooling run and we obtained equivalent results (at least in the regime of parameters relevant to this study and down to the scales comparable to few smallest resolution elements).\\

\subsection[]{AGN Heating}

The heating module consists of two key elements: computation of mass accretion onto the central 
supermassive black hole, and the energy injection via collimated bipolar outflows.
The quantity of accreted material is estimated as the
inflow through a spherical surface very near the center, $r\sim500$ pc.
Each cubical cell $j$
containing a piece $A_j$ of the spherical surface, contributes 
$\dot{M}_{{\rm acc}_j}=\rho_j v_{r_j} A_j$
to the accretion rate only if the radial velocity $v_{r_j}$ is negative (inflow).
The accreted gas, $\dot{M}_{{\rm acc}_j}\Delta t$, is then removed from the corresponding zone.

After computing the total mass accretion rate, we calculate 
the kinetic power of the bipolar outflows from
\begin{equation}\label{jet}
P_{\rm jet} = \epsilon\,\dot{M}_{\rm acc}\, c^2,
\end{equation}

\noindent
where $\epsilon$ is the (mechanical) efficiency of the feedback, which is a free parameter of our model.
We note that the actual accreted gas should be ideally calculated a few gravitational radii from the 
black hole. This would require an extreme dynamical range,
which is currently not possible due to hardware limitations. 
However, we can simply absorb any uncertainty in the actual accretion rate 
into our efficiency parameter $\epsilon$. If only a fraction $f$ of the gas 
entering our sink region falls into the black hole, 
the real efficiency would need to be rescaled to $\epsilon/f$.

The AGN mass, momentum, and energy are injected into the volume through the internal boundaries in the middle of the computational domain. The total cross-sectional area of the nozzle is $A_{\rm noz}=(2\Delta x)^2$, where $\Delta x$ is the highest resolution in the domain. 
Both jets originate at the same interface 
along the positive and negative $z$-direction (${\bf n}_z$ is the $z$-vector). The jet flux terms
appear on the right hand side of Equations (\ref{cont}), (\ref{mom}) and (\ref{ene}).
This method of including AGN feedback appears to be slightly
better than injecting quantities at zone centers,
which may lead to artificial evacuation of the regions perpendicular to the jets. 

The integrated mass outflow rate, $\dot{M}_{\rm jet}\equiv\dot{m}_{\rm jet} A_{\rm noz}$, 
is based on the conservation of momentum ($M_{\rm jet} {\bf v}_{\rm jet}$), leading to
\begin{equation}\label{Mout}
\dot{M}_{\rm jet} = 2\,P_{\rm jet}/v^2_{\rm jet},
\end{equation}
where $v_{\rm jet}$ is the subrelativistic 
outflow velocity fixed at $5\times10^4$ km s$^{-1}$. The velocity of AGN outflows
is still poorly constrained and observations suggest values from several $10^3$ to
several $10^4$ km s$^{-1}$ (e.g., \citealt{cre03,tom10}). 
Our choice of the outflow velocity gives reasonable values
for the mass outflow rate of $\sim M_\odot$ yr$^{-1}$. Notice that more massive (and slower)
outflows tend to carve a more visible channel in the ICM, while lighter (and faster) jets
release most of their energy in the central region, producing strong shocks and big bubbles
similar to X-ray data (see G11a,b).

We also implemented wobbling jets, with variable
inclination ($\theta$) and azimuthal angle ($\phi$). In these computations the location of the nozzle remains fixed in space, but all three velocity components
(e.g., $v_x=v_{\rm jet}\,{\rm sin}\theta\, {\rm cos}\phi$) vary with time. 
The injection is still
``cylindrical'', in the sense that the velocity vectors at the nozzle remain parallel.
We consider simple models where jet orientation varies randomly within a specified solid angle. These models are further discussed in Section 3.2.3.\\

\subsection[]{Initial Conditions}

\begin{table} \label{params}
\caption{Model parameters}
\begin{tabular}{@{}lcccccc}
\hline 
Model     & efficiency $\epsilon$ & min($t_{\rm cool}/t_{\rm ff}$)  &  Notes\\ 
\hline
 pure CF      & ---     &      7, 21      &   no AGN heating       \\
 r7-6em3    & $6\times10^{-3}$   &      7              &  along $\pm {\bf n_z}$ \\
 r7-6em3w  & $6\times10^{-3}$   &     7              &  random wobbling \\ 
 r7-6em3c  & $6\times10^{-3}$   &      7              &  convergence ($2\Delta x$) \\
 r7-1em2    & $10^{-2}$                &      7              &  along $\pm {\bf n_z}$ \\
 r21-6em3  & $6\times10^{-3}$   &     21             &  along $\pm {\bf n_z}$ \\
 r21-1em2  & $10^{-2}$                &     21             & along $\pm {\bf n_z}$ \\
\hline
\\
\\
\end{tabular}
\end{table} 

We model our initial conditions on Abell 1795, which is a typical massive relaxed
cool-core cluster characterized by the virial mass of $\sim10^{15}$ M$_\odot$.
In order to compute the initial hydrostatic gas density profile, 
we take the observed temperature fit (Fig.~\ref{fig:7_CFpure_phases_prof}, second column, black dashed line) and use
the analytical formula for a NFW dark matter gravitational acceleration. 
The normalization of the density profile is chosen such that the
gas fraction is $\approx$ 0.15 at the virial radius.
The obtained initial conditions correspond to the 
initial minimum ratio of $t_{\rm cool}/t_{\rm ff}\sim 7$, where the
free-fall time is given by\\
\begin{equation}\label{tff}
t_{\rm ff} = \left(\frac{2r}{g(r)}\right)^{1/2}.
\end{equation}

Our computational domain is
1.27 Mpc on a side and we use outflow boundary conditions.
We use static mesh refinement.
The linear size of each shell doubles among adjacent levels.
The highest refinement level is 10, and every block is divided into $8^3$ zones.
The finest zone of the box has a 
size of $\Delta x \sim 300$ pc.

As noted by S11, symmetric initial conditions do not produce thermal instabilities.
Therefore, we perturb the initial density and temperature profiles
to initiate the formation of cold filaments and blobs.
Depending on the model parameters, 
such features can develop even in the early stages of evolution, 
before the AGN has become very powerful.
We also note that the presence of the instabilities early on facilitates 
dispersal of the energy injected
by the collimated AGN jets. 

The fluctuations leading to the thermal instability can be produced by the turbulence
associated with cosmological mergers. It is beyond the scope of this
work to study such effects. It is thus sufficient to adopt a simple Gaussian random
field with a white noise power spectrum of gas density.
Using dispersion relation and direct simulation, M11 and S11 
demonstrated that thermal instability does not depend on the 
range of the driving modes in Fourier space. Moreover, in the late stage of the evolution,
the density fluctuations are dominated by the AGN jets. 

We generate initial conditions using IDL.
First, 
we initialize a complex stochastic field ($1024^3$) in the ${\bf k}$-space,
with amplitude of each mode given by
\begin{equation}\label{white}
W\propto k^{-1},
\end{equation}
where $k=(k^2_x+k^2_y+k^2_z)^{1/2}$. The above prescription ensures that
the `3D' power spectrum, which is $\propto W^2 \,k^2$, remains constant, i.e., is described by 
a white noise spectrum.
Second, we combine the above amplitude with a Gaussian random field (e.g., \citealt{rus07}):
\begin{equation}\label{gauss} 
[{\rm Re}(W({\bf k})),\; {\rm Im}(W({\bf k}))] = [G(u_1)W,\;G(u_2)W] 
\end{equation}
where $G$ is a function of a uniform random deviate $u_1$ or $u_2$ that
returns Gaussian-distributed values. This method guarantees that the phases
are random. We also impose cutoffs at 
$k_{\rm max}=1024/4$ and $k_{\rm min}=3$ (in units of $2\pi/L$). The high-k
cutoff is introduced in order to avoid spurious numerical effects.
Third, we compute inverse Fourier transform of $W({\bf k})$ to obtain $W(x,y,z)$.
We normalize it by the mean of its absolute value $\langle|W(x,y,z)|\rangle$.
Finally, we superpose the normalized fluctuations onto the initial density distribution
$\rho_{\rm unp}$:
\begin{equation}\label{pert}
\rho(x,y,z) = \rho_{\rm unp}(x,y,z)\left(1+\xi \frac{W(x,y,z)}{\langle|W(x,y,z)|\rangle}\right),
\end{equation}
where $\xi =0.3$ is the amplitude of the perturbations.
The new temperature distribution is computed from the new density assuming isobaric fluctuations.\\

\section[]{Results}
We present simulation results for different AGN efficiencies $\epsilon$, methods of energy injection (steady and wobbling jets), and different initial values of the minimum $t_{\rm cool}/t_{\rm ff}$ in the ICM (TI-ratio; minimum value over all radii). 
Our previous results (G11a; S11) demonstrated that a suitable choice of the mechanical AGN efficiency 
may be in the range $10^{-3}-10^{-2}$. This is supported by observational constraints (e.g., \citealt{meh08}).
Common observed values for the minimum initial TI-ratio are in the range from 5 to 25 (M11).
Motivated by these results, we perform computations with initial TI-ratio 7 or 21, and $\epsilon=$
$6\times10^{-3}$ or $10^{-2}$ (see Table 1).\\

\subsection[]{Pure Cooling Flow}

\begin{figure} 
    \subfigure{\includegraphics[scale=0.432]{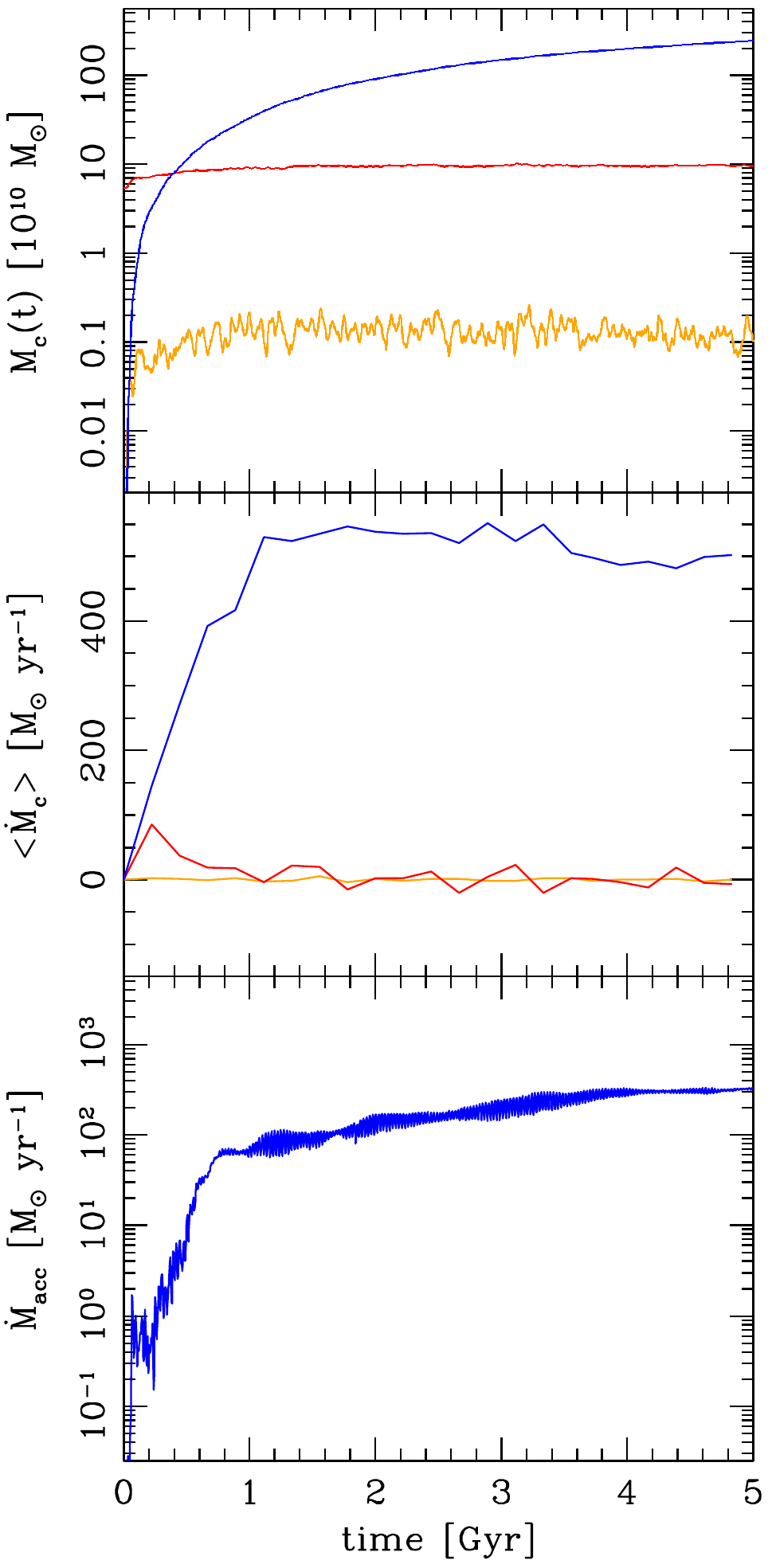}}
    \subfigure{\includegraphics[scale=0.432]{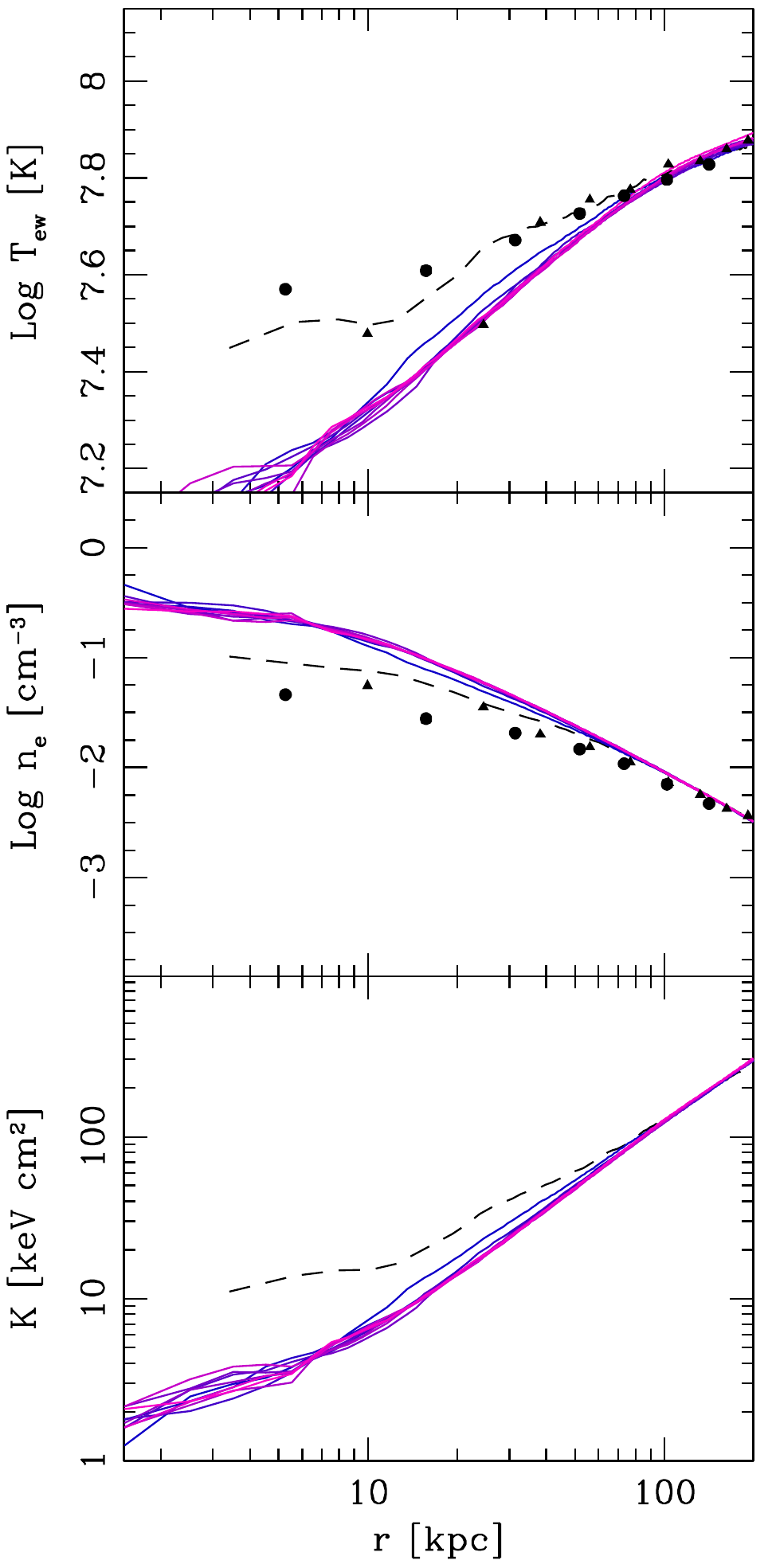}}
     \caption{Diagnostics for the pure cooling flow 
     simulation with initial minimum $t_{\rm cool}/t_{\rm ff}=7$. 
     First column (from top panel): total mass and average mass cooling rate as a function of time in the central 20 kpc; accretion rate (in the sink region) as a function of time. Colors correspond to the three gas phases:
     blue - cold ($T<5\times10^5$ K), yellow - warm ($5\times10^5\le T<10^7$ K), 
     red - hot ($T\ge10^7$ K). Notice the strong cooling rate of $\sim500$ $\msun$ yr$^{-1}$.
           Second column: radial profiles of the emission-weighted temperature, electron number density, and entropy ($k_{\rm b}T/n_{\rm e}^{2/3}$).
          Curves are plotted every 500 Myr, and the color changes gradually from dark blue to light violet; black dashed lines correspond to initial conditions, which are based on observational data of Abell 1795 (circles, \citealt{tam01}, and triangles, \citealt{ett02}).
     \\
     \\
     \label{fig:7_CFpure_phases_prof}}
\end{figure}

\begin{figure*}  
        \subfigure{\includegraphics[scale=0.31]{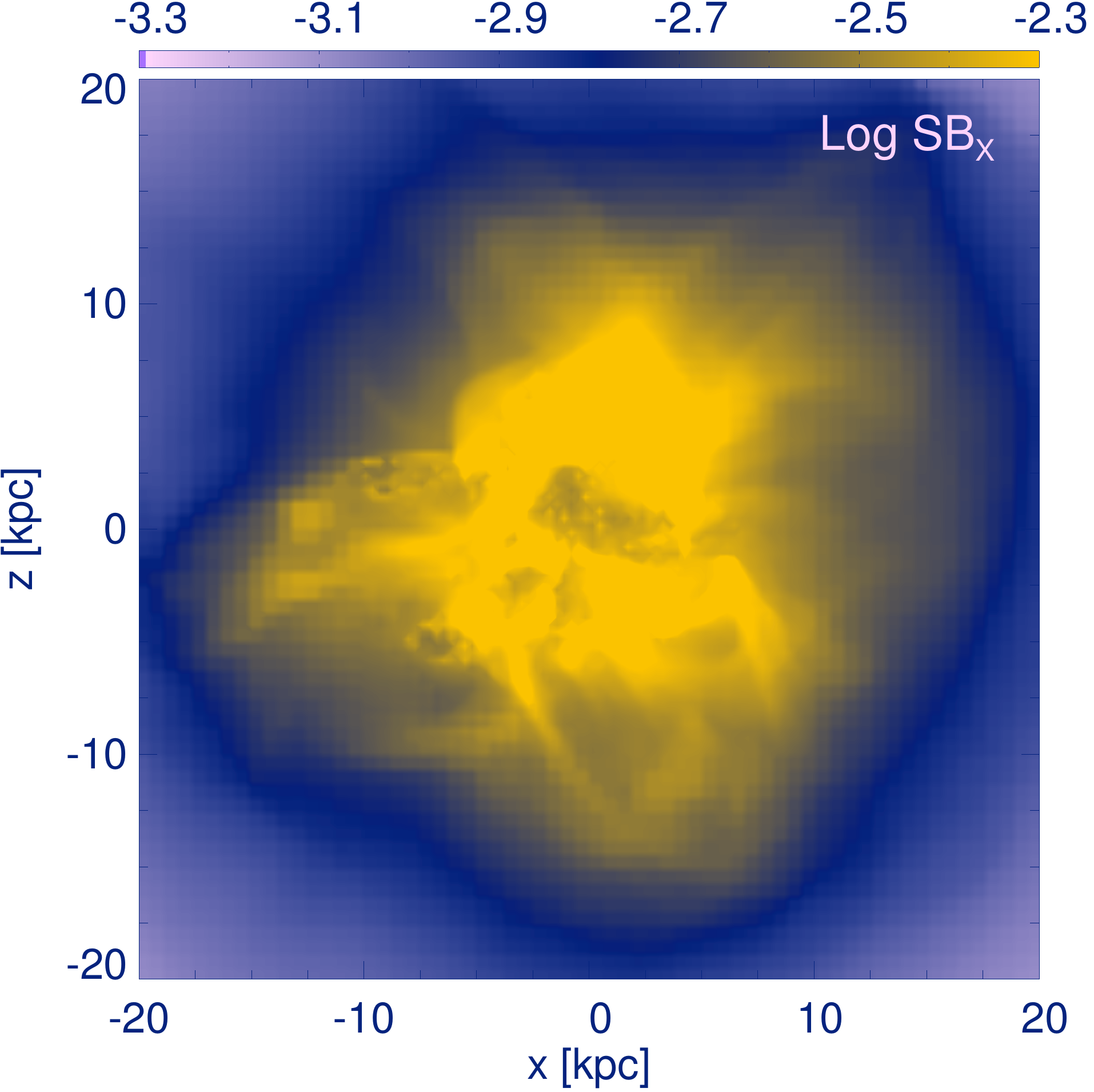}}
    \subfigure{\includegraphics[scale=0.31]{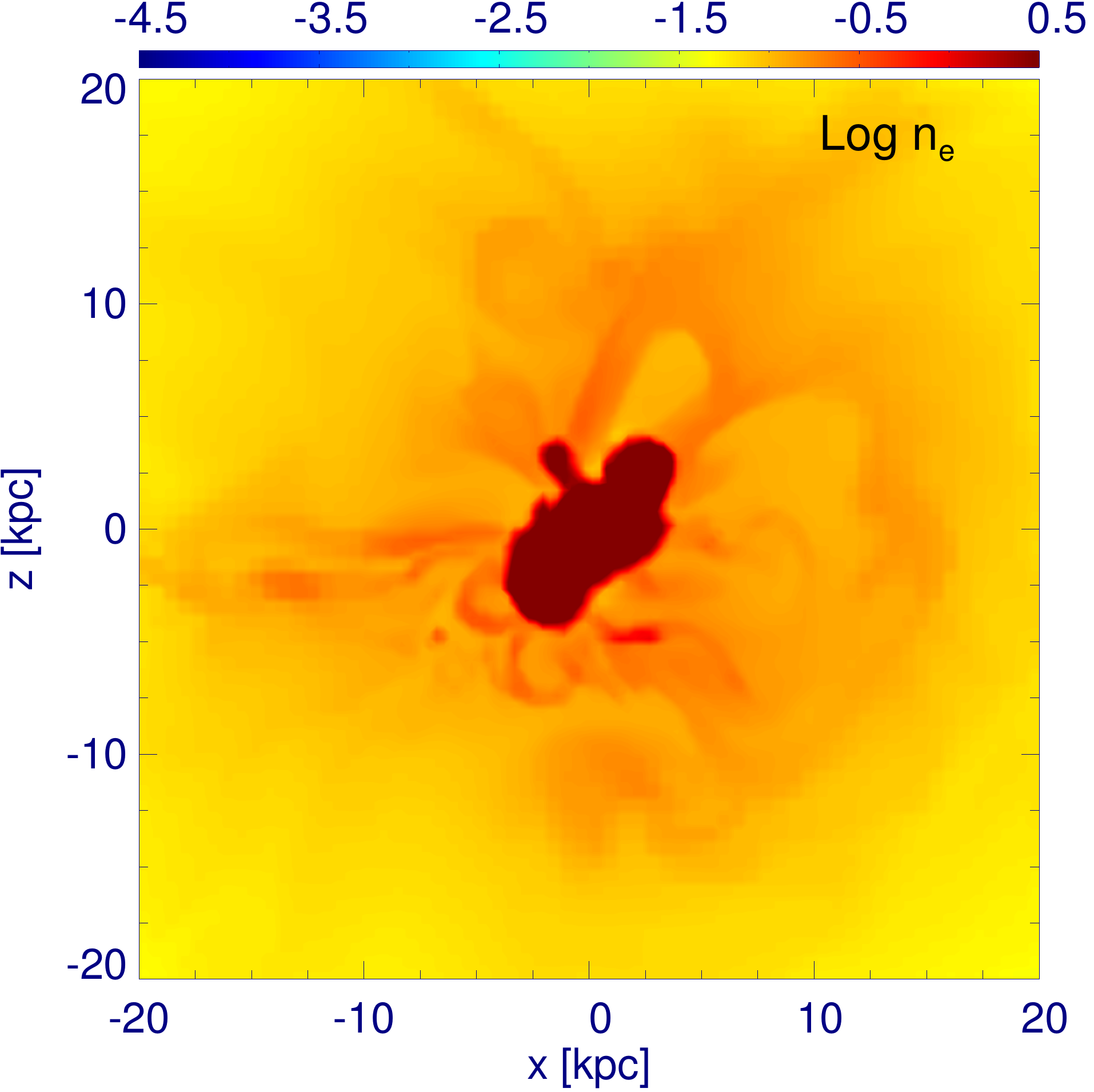}}
    \subfigure{\includegraphics[scale=0.31]{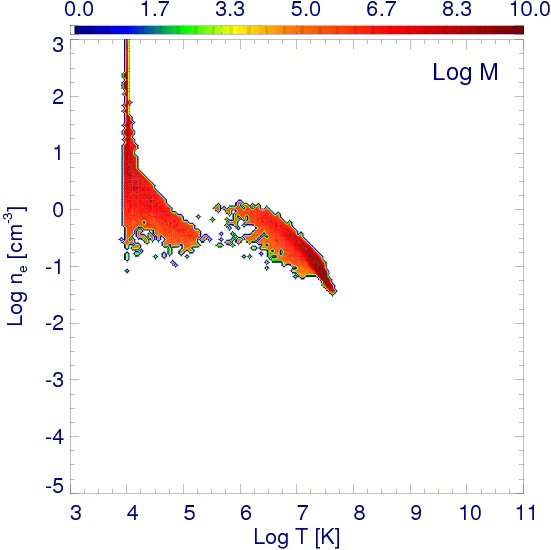}}    
    \caption{Pure CF simulation at 1.51 Gyr. Left: 
     X-ray surface brightness (erg s$^{-1}$ cm$^{-2}$).
     Middle: $n_{\rm e}$ midplane cut (cm$^{-3}$).
     Right: joint distribution of $\log T$ (abscissa) versus $\log n_{\rm e}$ (ordinate), colored by gas mass in each bin ($\msun$); the size of the logarithmic bins is 0.05.
     Each map has logarithmic scale (see color bars) and, apart from the joint distribution, is 20 kpc on a side (i.e., only the very central regions of the entire computational domain are shown).
     Thermal instabilities and extended cold gas are possible only in the early stage; later
     the collapse becomes monolithic, with a huge central accumulation of cold gas.
     \\
     \\
     \label{fig:7_CFpure_maps}}
\end{figure*}

We first analyze a reference model, i.e., a simulation without AGN heating. In this simulation
we keep the central gas sink term but neglect feedback. 
This case leads to the classic `cooling flow' catastrophe.
The plasma in the central region of the cluster ($\sim$100-200 kpc) 
looses energy due to radiative cooling. Consequently, the gas pressure drops, leading
to a subsonic spherical inflow. The increase in the central gas density results in higher cooling rates,
which, in turn, accelerates the mass accretion rate, leading to runaway cooling and accretion.
The results from a cooling flow run for the initial TI-ratio of 7 are presented in Figures \ref{fig:7_CFpure_phases_prof} and \ref{fig:7_CFpure_maps}. 

In the first column of Figure \ref{fig:7_CFpure_phases_prof} we show the total mass, the average mass 
rate and the accretion rate as a function of time.
The quantities in the first two panels correspond to the gas within the central 20 kpc. Different gas phases are coded according to: `cold' ($T<5\times10^5$ K, blue), `warm' ($5\times10^5\le T<10^7$ K, yellow), `hot' ($T\ge10^7$ K, red). The choice is not arbitrary because the limits of these temperature ranges correspond approximately to different critical slopes in the cooling function $\Lambda(T)$. 
Top panel reveals that the total mass of cold gas exceeds $10^{12}$ $\msun$ after a few Gyr.
Middle panel shows that after 1 Gyr the cold phase continues to
accumulate in the core (20 kpc) at a steady average rate of $\sim500$ $\msun$ yr$^{-1}$ (peaks can reach $\sim800$ $\msun$ yr$^{-1}$), a blatant discrepancy with observations. The cold gas never decreases with time, usually overwhelming the hot and warm phase by one and three orders of magnitudes, respectively.
The accretion rate in the sink region 
is closely linked to the cooling rate. However, not all of the cooling gas gets accreted
because the inflowing cold gas forms a rotationally-supported structure.

The second column displays the radial profiles of the emission-weighted temperature, electron number density, and entropy (for the X-ray emitting gas with $T>0.3$ keV).
Curves are plotted every 500 Myr, and the color changes gradually from dark blue to light violet; black dashed lines correspond to the initial conditions. 
The pure cooling flow manifests itself as a sudden decrease in temperature, lesser than $T_{\rm vir}/3$.
The projected emission-weighted ($T_{\rm ew}$) temperatures
falls in particular below the observational data of Abell 1795 (circles, \citealt{tam01}, and triangles, \citealt{ett02}). The electron number density $n_{\rm e}$, linked to the surface brightness, increases rapidly within 70 kpc from the center, with the simulated profiles lying above the observational data points.
The entropy ($k_{\rm b}T/n_{\rm e}^{2/3}$) near 1 kpc is $\sim$ 1 keV cm$^2$,  
without showing any sign of the typical observed floor (\citealt{cav08}).

In Figure \ref{fig:7_CFpure_maps}, we present the maps of
X-ray surface brightness, 
the midplane cut of electron number density ($n_{\rm e}$), and the gas
phase diagrams with temperature versus $n_{\rm e}$ (values are color-coded by gas mass in each bin).
Each map has logarithmic scale and contains the data only for the central core (20 kpc);
the snapshot is taken at 1.51 Gyr. 
The density maps reveal significant cold gas accumulation in the core, accompanied by a strong increase in the
central surface brightness.
Since in the `warm' temperature range the gas has a short cooling time,
the ICM tends to accumulate in the `cold' and `hot' phases.
This effect is clearly visible in the phase diagram. 


In the cooling flow run distinct cold gas clouds and filaments can be seen predominantly in the early stages of the evolution. 
At late times (e.g., 1.51 Gyr) the collapse proceeds in a monolithic way, and the atmosphere is globally unstable.
Note that the $t_{\rm cool}/t_{\rm ff}$ criterion does not apply to a cooling flow, because the theoretical analysis assumes thermal balance (see M11). In order to generate multiphase gas in a cooling flow, a substantial amplitude of perturbations
is required, e.g., $\gta0.1$ (see also \citealt{pis05}; S11). Both initial TI-ratio of 7 and 21 (not shown) produce similar results. The latter 
case has longer saturation time and slightly smaller mass in the cold phase 
(the initial total mass in the box
is also 8\% smaller). 

All the features of the pure CF model, discussed above, are in contrast with observations of galaxy clusters (\citealt{pet01,pet03,ett02,tam03,pef06}). This simulation merely serves as a reference case for the runs presented in the subsequent sections. A successful model will instead be characterized by a
quasi equilibrium between heating and cooling, with low cooling rates ($\la 10\%$ of the pure
cooing flow), profiles consistent with observations
(e.g., emission-weighted radial profiles with reasonable gradients and characteristic length scales). At the same time, 
the model needs to account for the spatially extended (up to $\sim15-20$ kpc) distribution of cold blobs and filaments while not violating the constraints on the total mass in these features. Below we discuss models that meet these requirements.\\

\subsection[]{AGN Feedback: $t_{\rm cool}/t_{\rm ff}=7$}

Before including AGN jet feedback, we verified that using the idealized prescription
of S11, i.e., local heating rate given by the cooling rate averaged in
spherical shells, produces thermal instabilities and spatially extended multiphase gas. The results were consistent with S11.
However, unless cooling and heating are turned off in the central zones (as in M11),
this simple prescription led to the massive accumulation of cold gas in the central few kpc,
especially when the gas dropout or outflow through a central boundary was neglected. 
These findings provide additional motivation to include a realistic feedback 
mechanism. Below we discuss the results of our simulations that include self-regulated feedback due to collimated AGN jets.

\subsubsection[]{$\epsilon=6\times10^{-3}$}

\begin{figure*} 
    \subfigure{\includegraphics[scale=0.453]{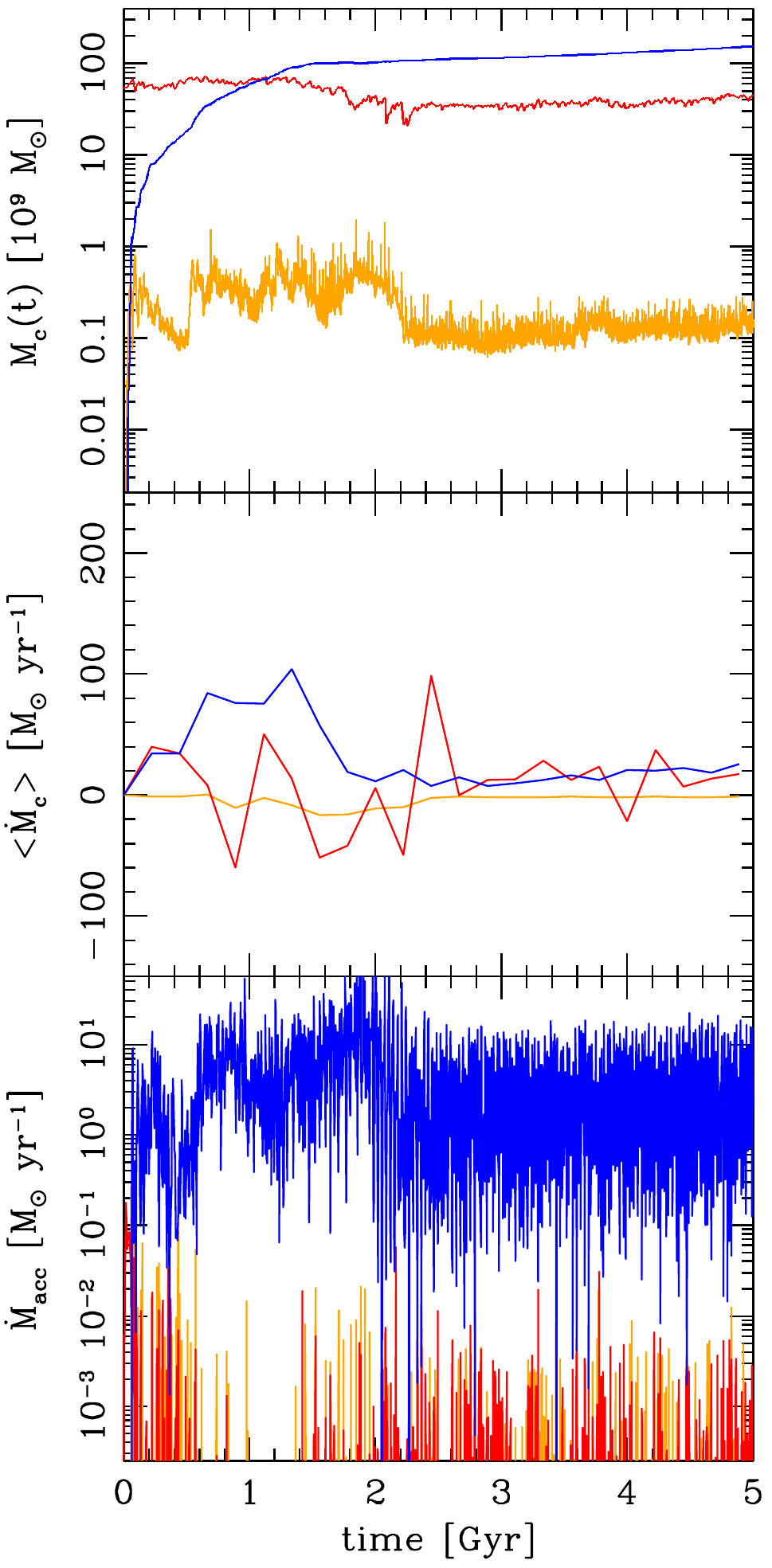}}
    \subfigure{\includegraphics[scale=0.453]{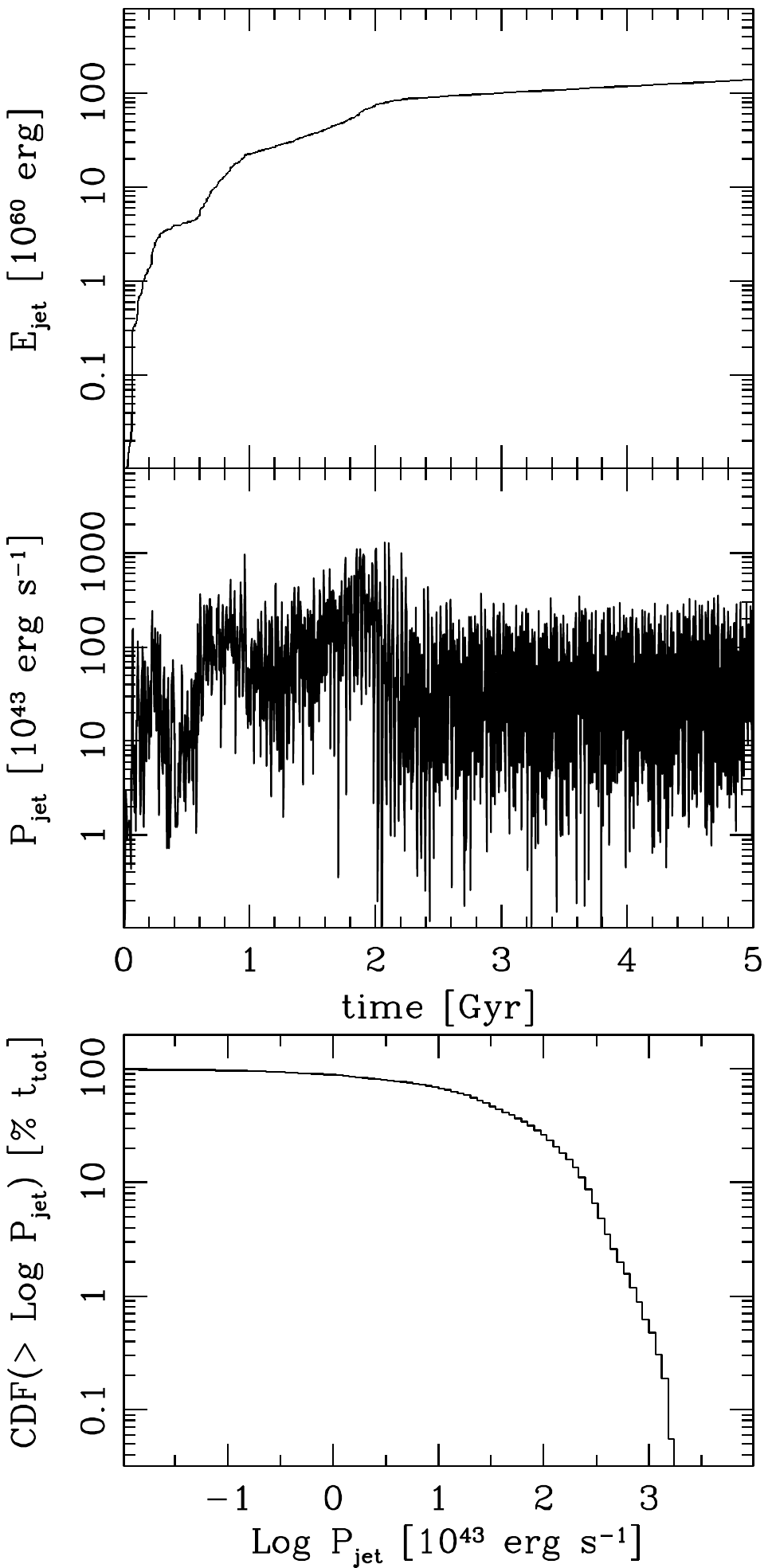}}
    \subfigure{\includegraphics[scale=0.453]{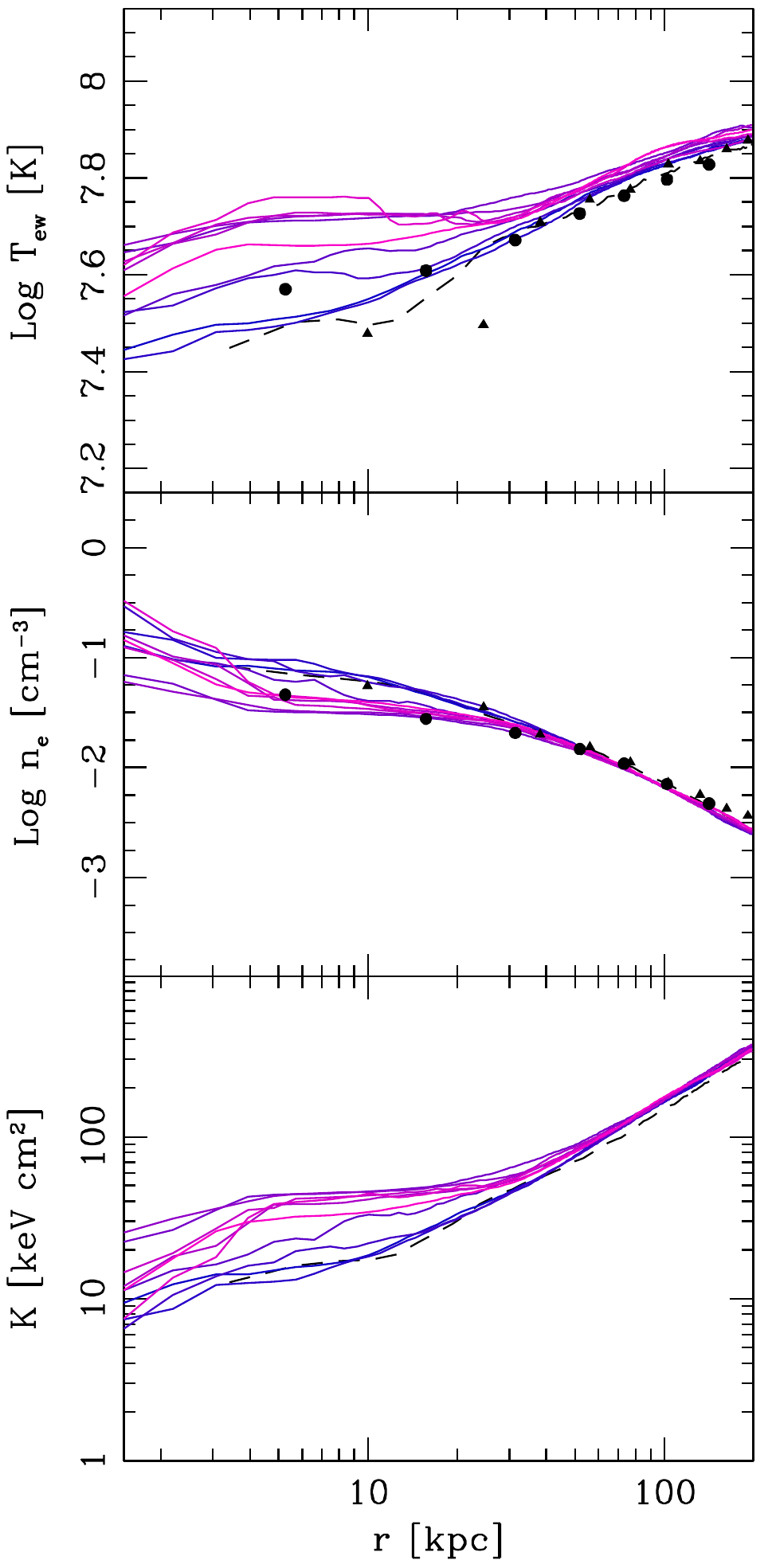}}
    \subfigure{\includegraphics[scale=0.453]{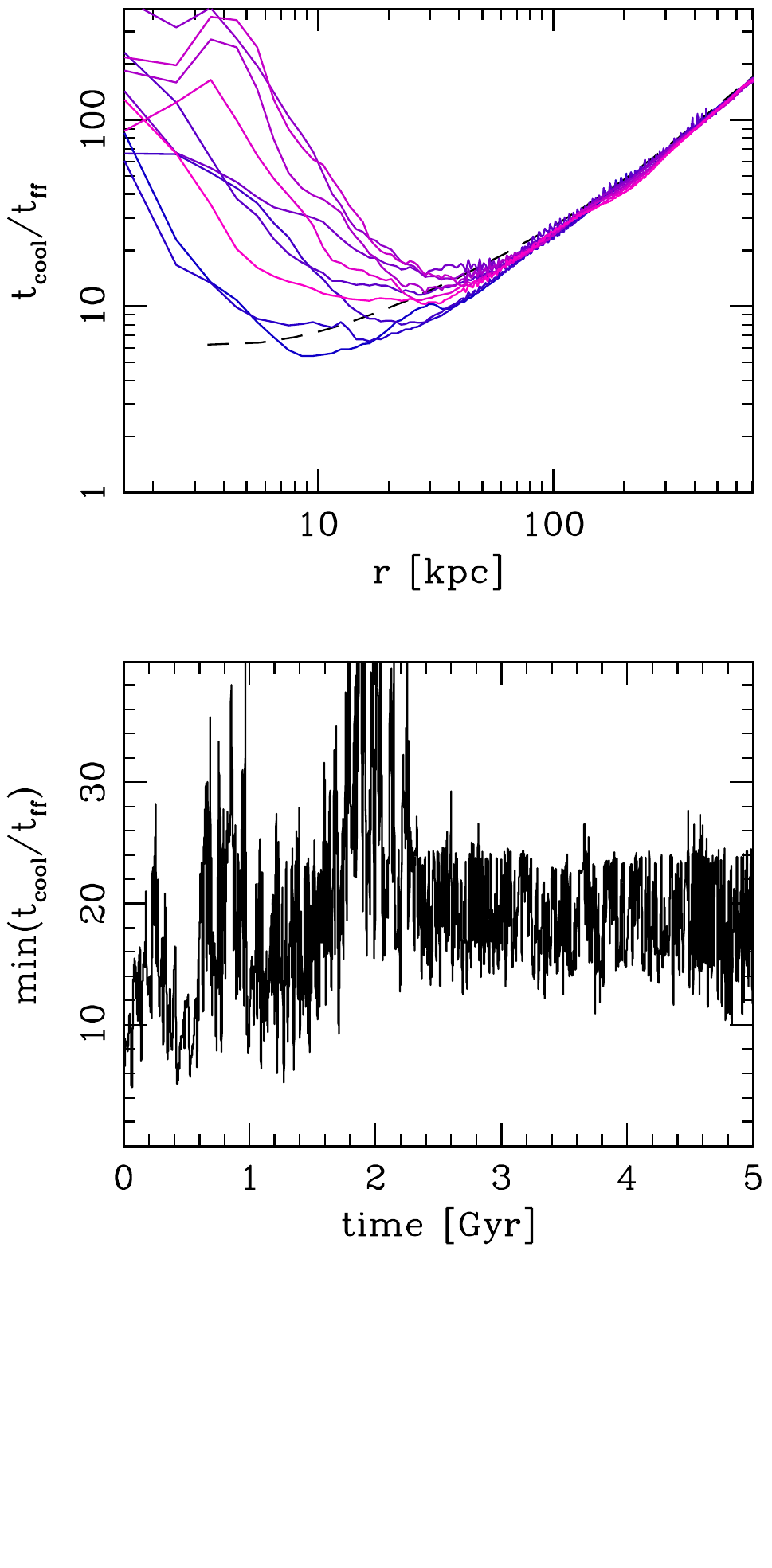}}    
     \caption{Diagnostics for the AGN feedback simulation with initial minimum $t_{\rm cool}/t_{\rm ff}=7$ and 
     $\epsilon=6\times10^{-3}$ (r7-6em3). 
     First column: cooling and accretion diagnostics; see the description of
     Figure \ref{fig:7_CFpure_phases_prof} for details.
     Second column (from top panel): total injected mechanical energy, 
     outflow power and cumulative distribution function (`duty cycle') of the jet powers (100\% corresponds to 5 Gyr); only for one of the jets.  
          Third column: radial profiles; see the description of
     Figure \ref{fig:7_CFpure_phases_prof} for details.
     Fourth column: radial profile of $t_{\rm cool}/t_{\rm ff}$ and the temporal evolution of its minimum (gas with $T>0.3$ keV).
     Note that the cooling rate is reduced to $\sim4\%$ of the pure CF value and that
     a steady average equilibrium is maintained for several Gyr. 
     \label{fig:7_6em3_phases_jet} }
\end{figure*}
\begin{figure*}  
    \subfigure{\includegraphics[scale=0.31]{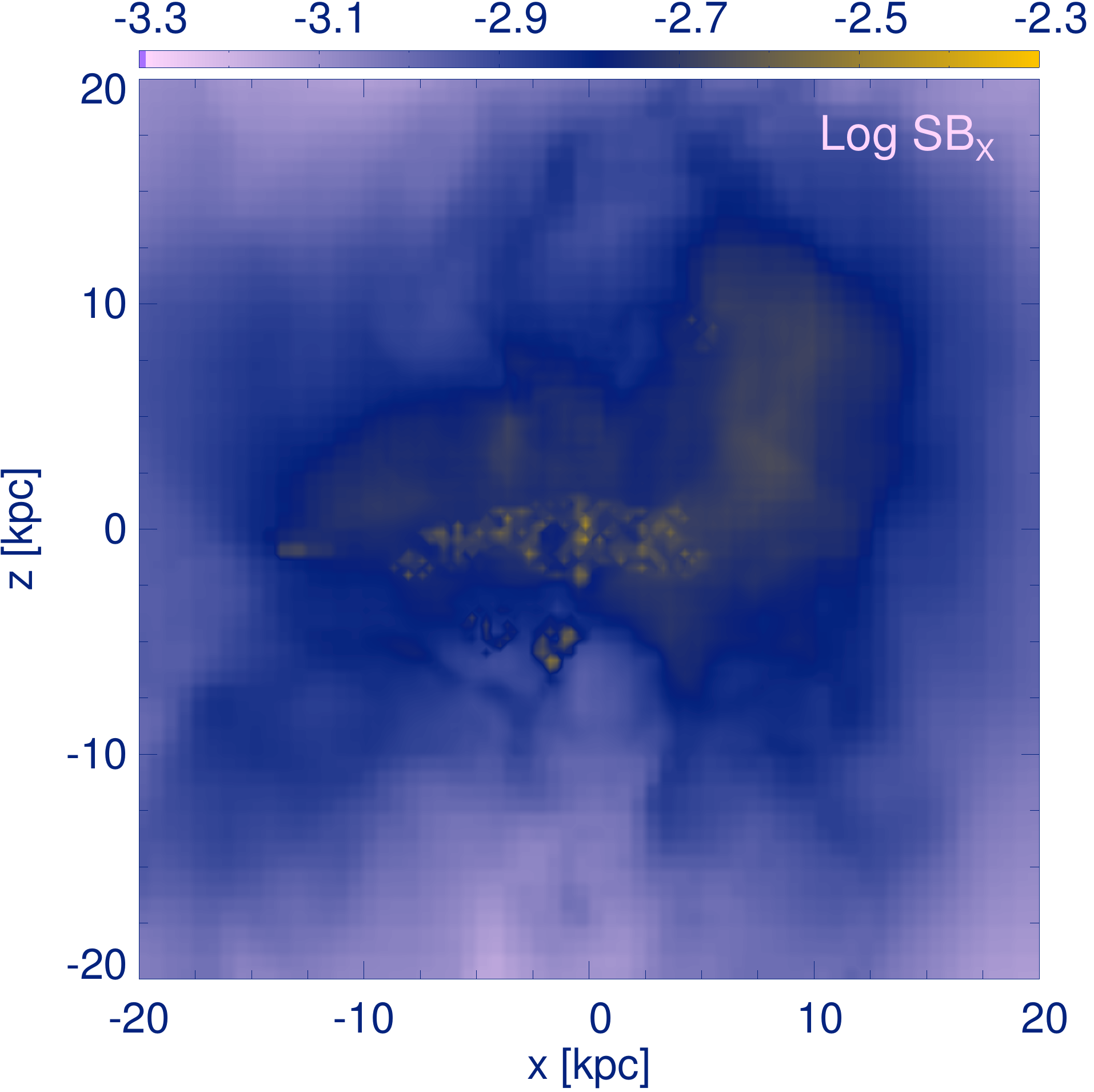}}
    \subfigure{\includegraphics[scale=0.31]{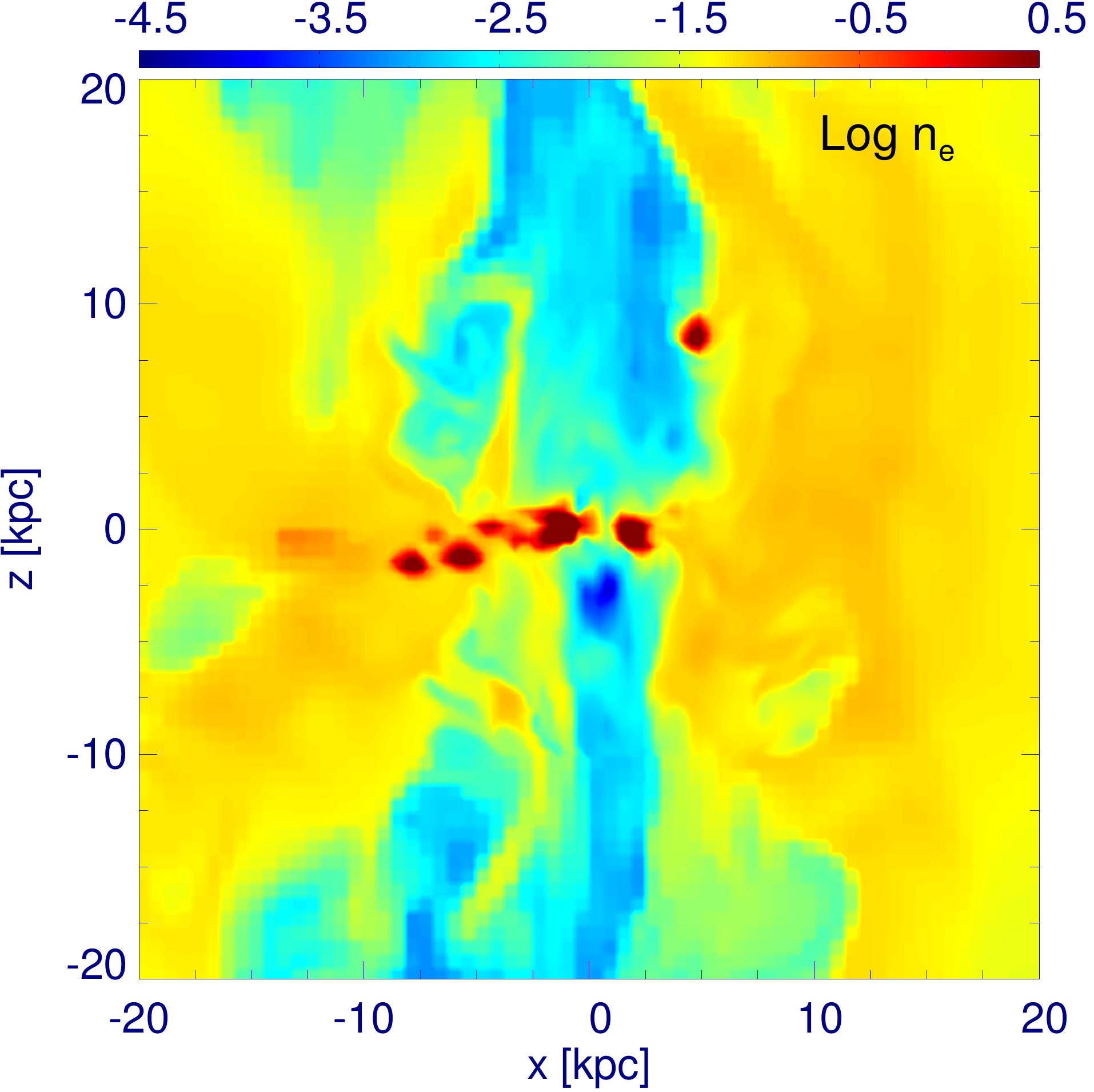}}
    \subfigure{\includegraphics[scale=0.31]{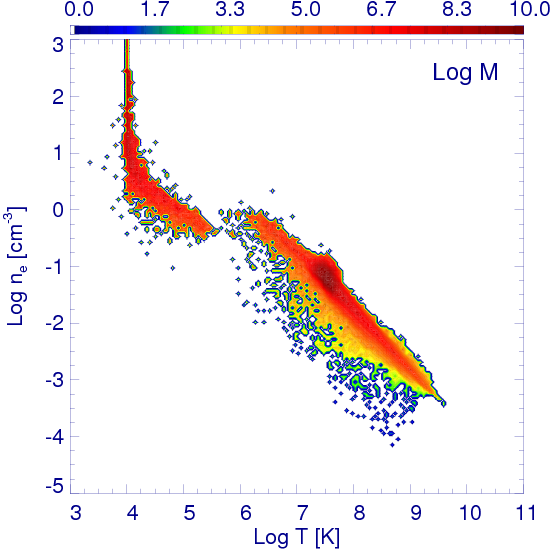}}
     \caption{Maps for the same simulation as in Fig.~\ref{fig:7_6em3_phases_jet} (r7-6em3). See the description of
     Figure \ref{fig:7_CFpure_maps} for details.
          The maps correspond to the time of 1.51 Gyr since the beginning of the simulation. 
     The salient feature is the amplification of thermal instabilities, due to the AGN perturbations,
     with cold blobs condensing out of the hot phase ($t_{\rm cool}/t_{\rm ff}\la10$) 
     up to 10 kpc away from the center. 
     \\
     \\
     \label{fig:7_6em3_maps} }
\end{figure*} 

Mechanical feedback due to AGN jets dramatically changes the entire evolution of the galaxy cluster, compared to the pure cooling case (Sec.~3.1).  In Figures \ref{fig:7_6em3_phases_jet} and \ref{fig:7_6em3_maps} we present diagnostic plots and maps for the runs that include AGN jets with mechanical efficiency $\epsilon=6\times10^{-3}$.

The first column in Fig.~\ref{fig:7_6em3_phases_jet} show the same quantities as those shown in the 
left panels in Fig.~\ref{fig:7_CFpure_phases_prof}. The gas mass in the cold phase $M_{\rm c}(t)$, shown in the top panel, reaches $\sim10^{11}$ $\msun$ at the end of the evolution. This is consistent with molecular cold gas observations (\citealt{edg01,sal03}), and is reduced by over a factor of 10 compared to the pure cooling flow value. Middle panel shows that the average cooling rate $\langle\dot{M}_{\rm c}\rangle$ is also significantly quenched: after an initial transition phase, in which cooling is slightly elevated, the system starts to self-regulate through more powerful outbursts, and eventually reaches a steady rate of $\sim20$ $\msun$ yr$^{-1}$. This value is about $4\%$ of the value seen in the pure CF simulation, entirely consistent with observations (e.g., \citealt{pef06}).
We note that the instantaneous cooling rate (not shown) is quite variable, changing rapidly from positive to negative values due to the action of AGN heating. 

As the cold gas mass $M_{\rm c}$ in the central 20 kpc exceeds the mass in the hot (3 times) and warm phases
($10^3$ times), the accretion rate is determined by the cold gas (last panel of first column);
most of the time the black hole accretes cold gas at a rate between $1-5$ $\msun$ yr$^{-1}$.
Cold accretion rate (blue) 
exceeds hot accretion rate (red) 
by at least two orders of magnitude. The latter is commonly associated with a Bondi-like mode;
since the hot accretion rate is so weak, feedback simulations based on Bondi prescription usually require an artificial boost factor or high mechanical efficiencies $\gtrsim0.1$,
 even if the Bondi radius is resolved (e.g., G11a, \citealt{dim08}).
 However, we warn against interpreting our cold gas accretion rates literally, as the actual accretion rate onto the black hole will likely be lower.

The mass accretion rate determines the AGN power and mass outflow rate (see Eq.~(\ref{jet}) and (\ref{Mout})). The properties of one of the jets are shown
in the second column in Figure \ref{fig:7_6em3_maps}. The cumulative energy injection as a function of time is presented 
in the top panel. At 5 Gyr the total injected energy is $\sim10^{62}$ erg. Even if the entire mass of the gas that falls into the sink region is added to the SMBH, its total mass will still be within reasonable limits
(less than several $10^9$ $\msun$).
Typical values of the jet power (second panel) and mass outflow rate are $P_{\rm jet}\sim3-4\times10^{44}$ erg s$^{-1}$
and $\dot{M}_{\rm jet}\sim0.5-1$ $\msun$ yr$^{-1}$. 
The cumulative distribution function 
(CDF\footnote{CDF($>X$) represents the probability that the independent variable $x$ is greater than value $X$; 
100\% value corresponds to 5 Gyr ($t_{\rm tot}$).}
; last panel) of the jet power suggests that the feedback is characterized by a `duty cycle'. 
For example, assuming that the AGN is considered `active' only when the mechanical power exceeds
$10^{44}$ erg s$^{-1}$, the duty cycle would be about $70\%$ (i.e., AGN active 70\% of the time). This quantity may be poorly defined because of its sensitivity to the chosen threshold. In particular, deeper observations may reveal more signs of AGN activity, leading to higher duty cycles.
Current observational estimates (e.g., \citealt{dun06}) place AGN duty cycle at a level $\gtrsim 70\%$.

The radial profiles (Figure \ref{fig:7_6em3_phases_jet}, third column) of the emission weighted temperature, electron number density, and 
entropy reveal the presence
of the feedback heating in two ways. First, the density inside 70 kpc is similar
to the initial (observed) conditions, meaning that the inflow due to the cooling is substantially quenched.
 Second, the temperature and entropy profiles do not fall below $T_{\rm vir}/3$ and 10 keV cm$^{2}$, respectively. The profiles 
maintain a positive gradient, a sign that the cool core has not been destroyed by the outbursts.
Notice that the real comparison with X-ray observations
must be based on emission-weighted temperatures: the pure mass-weighted $T$ profile would appear much more disturbed with higher central values. The electron number density profile ($n_{\rm e}$), also linked to the X-ray emissivity, appears consistent with observational data, with a flatter gradient compared to the pure CF run. 

In Fig.~\ref{fig:7_6em3_phases_jet} (third column) we show the radial profile 
of $t_{\rm cool}/t_{\rm ff}$ and its minimum as a function of time.
Although we discuss the relationship between $t_{\rm cool}/t_{\rm ff}$ and the presence of
the multiphase gas in more detail in Section 4, here we briefly point out that 
whenever this timescale ratio falls below $\approx 10$, 
thermal instabilities grow and cold gas clumps condense out of the hot phase. Initially, the cold phase has a filamentary structure; then it turns into small clumps which fall toward the center in about a free-fall time.

The SB$_{\rm X}$ map and the the two-dimensional snapshot of $n_{\rm e}$ at 1.5 Gyr, shown in Figure \ref{fig:7_6em3_maps} (first and second panel), corresponds to a moment when the TI-ratio falls below 10. 
The jet outburst creates two cavities along the $z-$direction,
while at the same time the perturbations due to the jets seem to lead to the formation of dense cold blobs 
up to $\sim 10$ kpc away from the center. 
The presence of multiphase gas is also apparent in the $n_{\rm e}-T$ phase diagram (Fig.~\ref{fig:7_6em3_maps}, third panel), which displays two distinct diagonal islands occupied by the gas in the cluster core, with small variations 
away from pressure equilibrium.

Because the AGN heating is anisotropic close to the base of the jet
and the cooling gas possesses small amount of angular momentum, 
the cold gas tends to accrete and condense more along the 
equatorial region. In a realistic situation this gas 
cannot be entirely removed in a brief time (as in S11); 
consequently, a cold rotationally-supported torus is formed, which provides most of the fuel for the AGN. The fuel supply from the torus, or the torus itself, can be temporarily disturbed by 
AGN outbursts, shocks, and the ICM turbulence. The central accumulation of cold gas could be further reduced by including star formation.

Overall, the heating due to realistic AGN jets, with mechanical efficiency $\approx6\times10^{-3}$, appears to reproduce the main properties of the observed clusters: quenching the cooling flows while preserving the cool cores for several Gyr.
At the same time, the model predicts the formation of the multiphase medium in the form of cold blobs and filaments while satisfying the observational constraints on the amount of cold gas
(see \citealt{mcd10,mcd11a,mcd11b}). 

\subsubsection[]{$\epsilon=10^{-2}$}

\begin{figure*} 
    \subfigure{\includegraphics[scale=0.453]{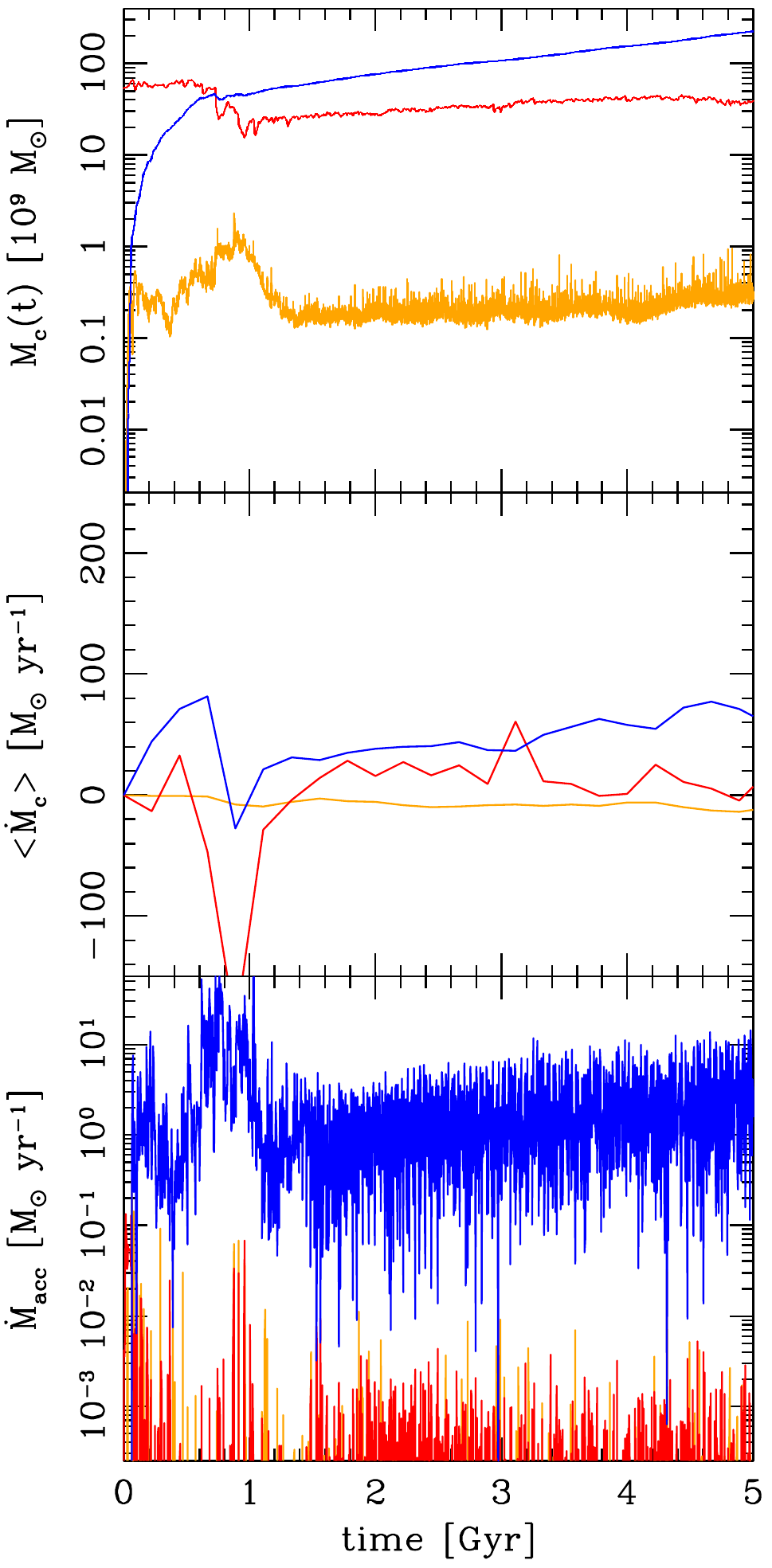}}
    \subfigure{\includegraphics[scale=0.453]{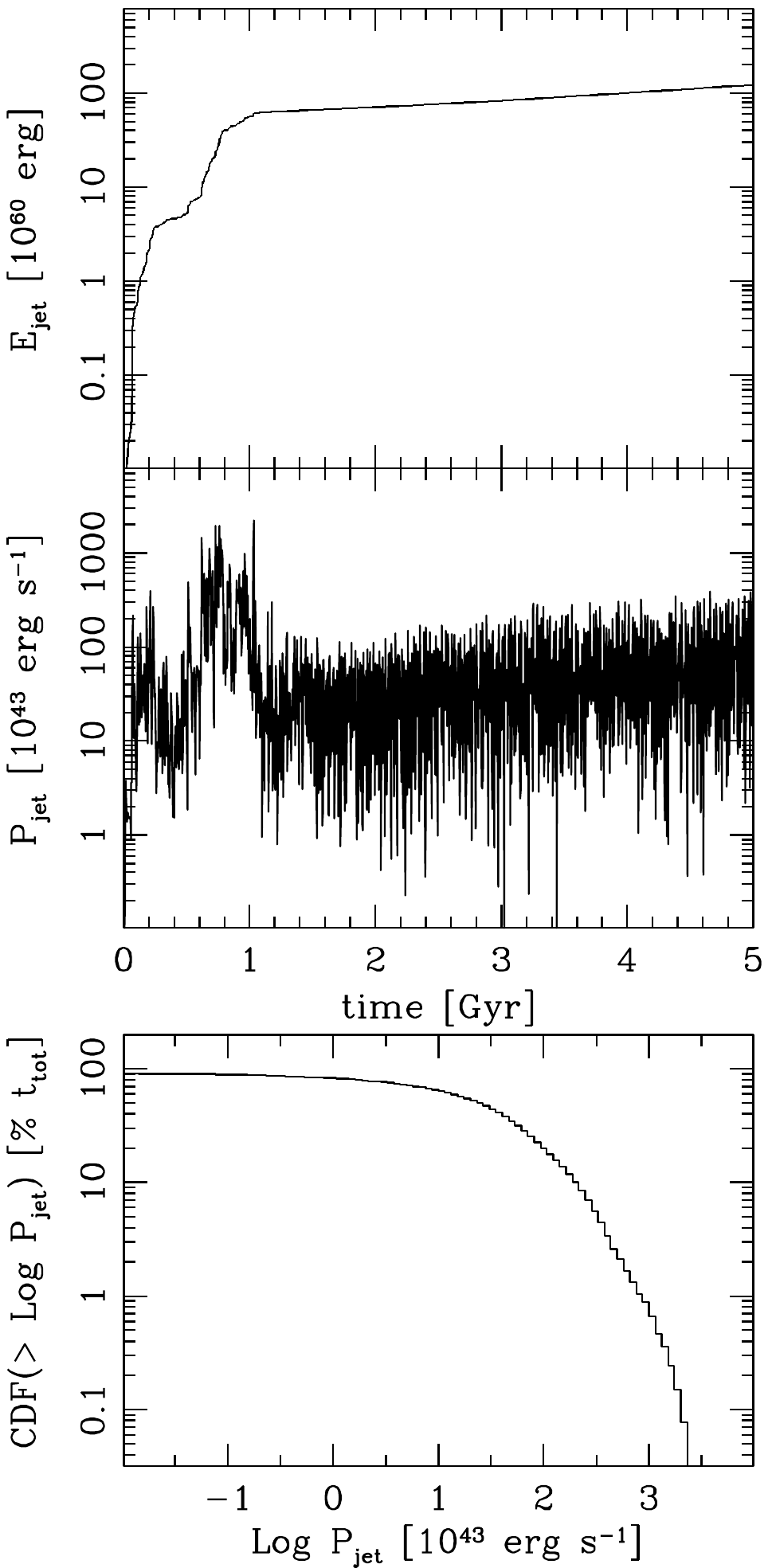}}
    \subfigure{\includegraphics[scale=0.453]{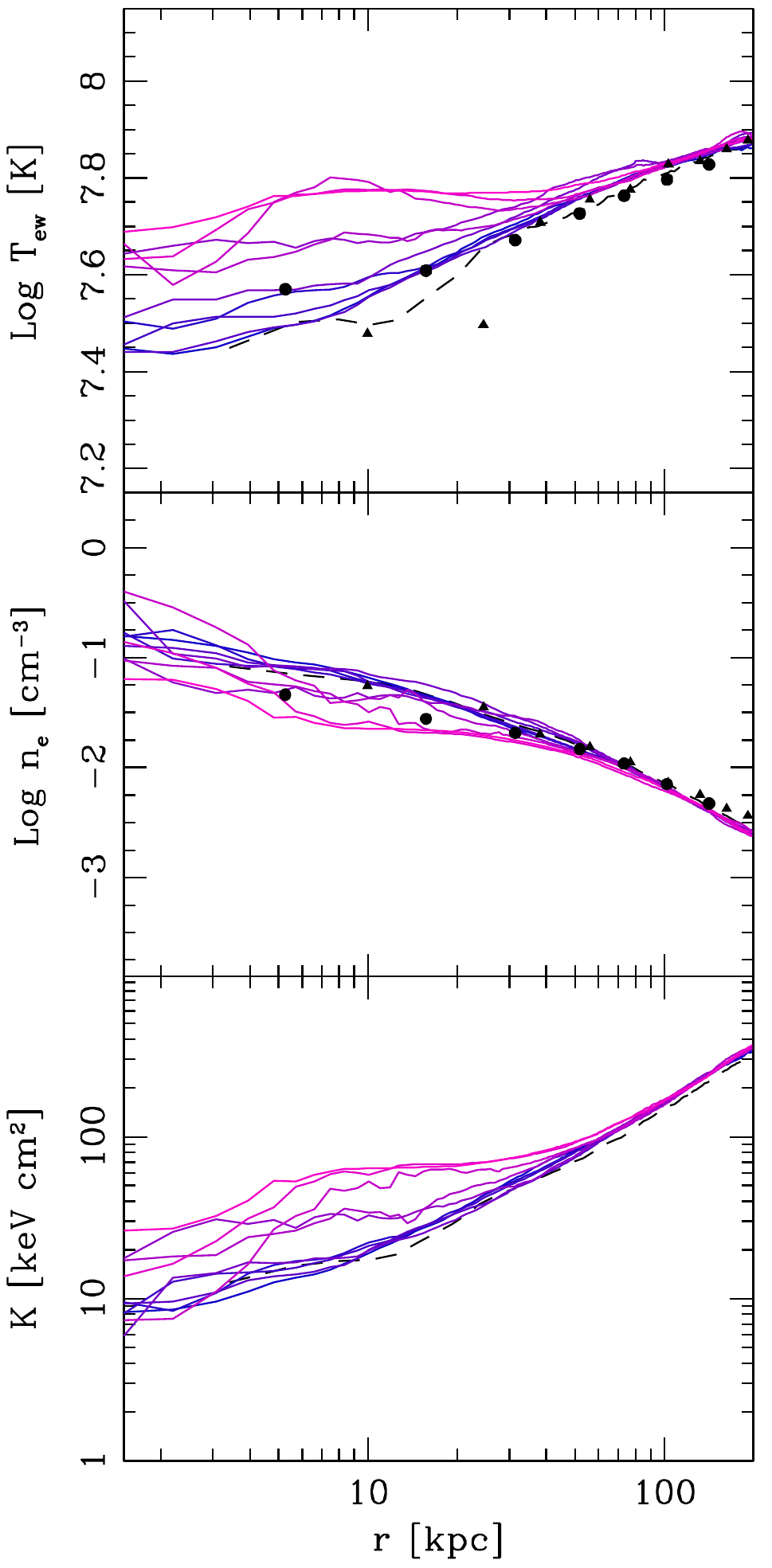}}
    \subfigure{\includegraphics[scale=0.453]{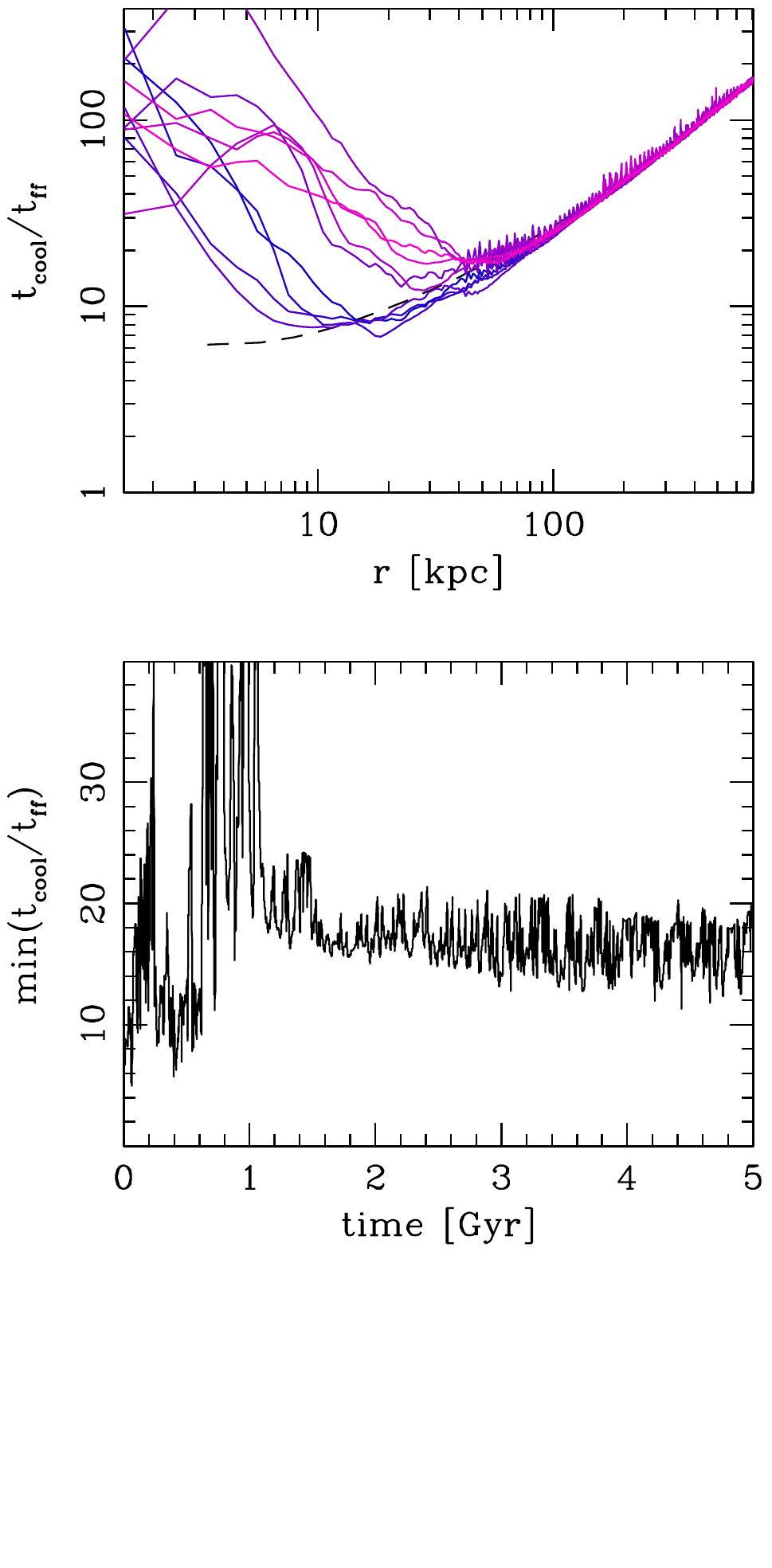}} 
     \caption{Same as Figure \ref{fig:7_6em3_phases_jet} but for AGN feedback efficiency $\epsilon=10^{-2}$ (r7-1em2).
          Raising the efficiency produces stronger and
     more massive outflows, which have more difficulty in stopping the cooling flow, because they
     tend to release their energy at larger radii. Values are still consistent with observations. 
     \label{fig:7_1em2_phases_jet}}
\end{figure*} 
\begin{figure*}  
    \subfigure{\includegraphics[scale=0.31]{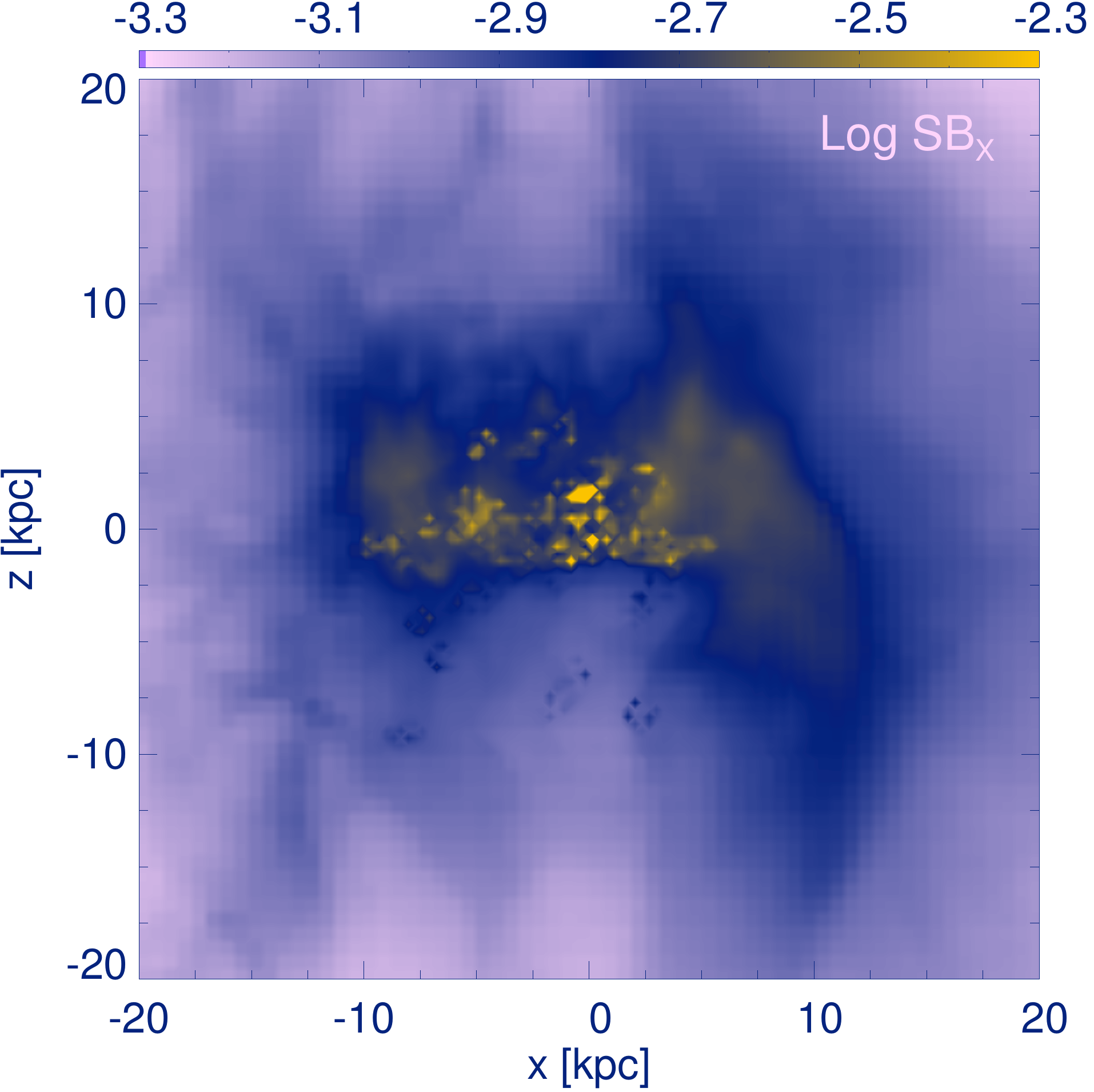}}
    \subfigure{\includegraphics[scale=0.31]{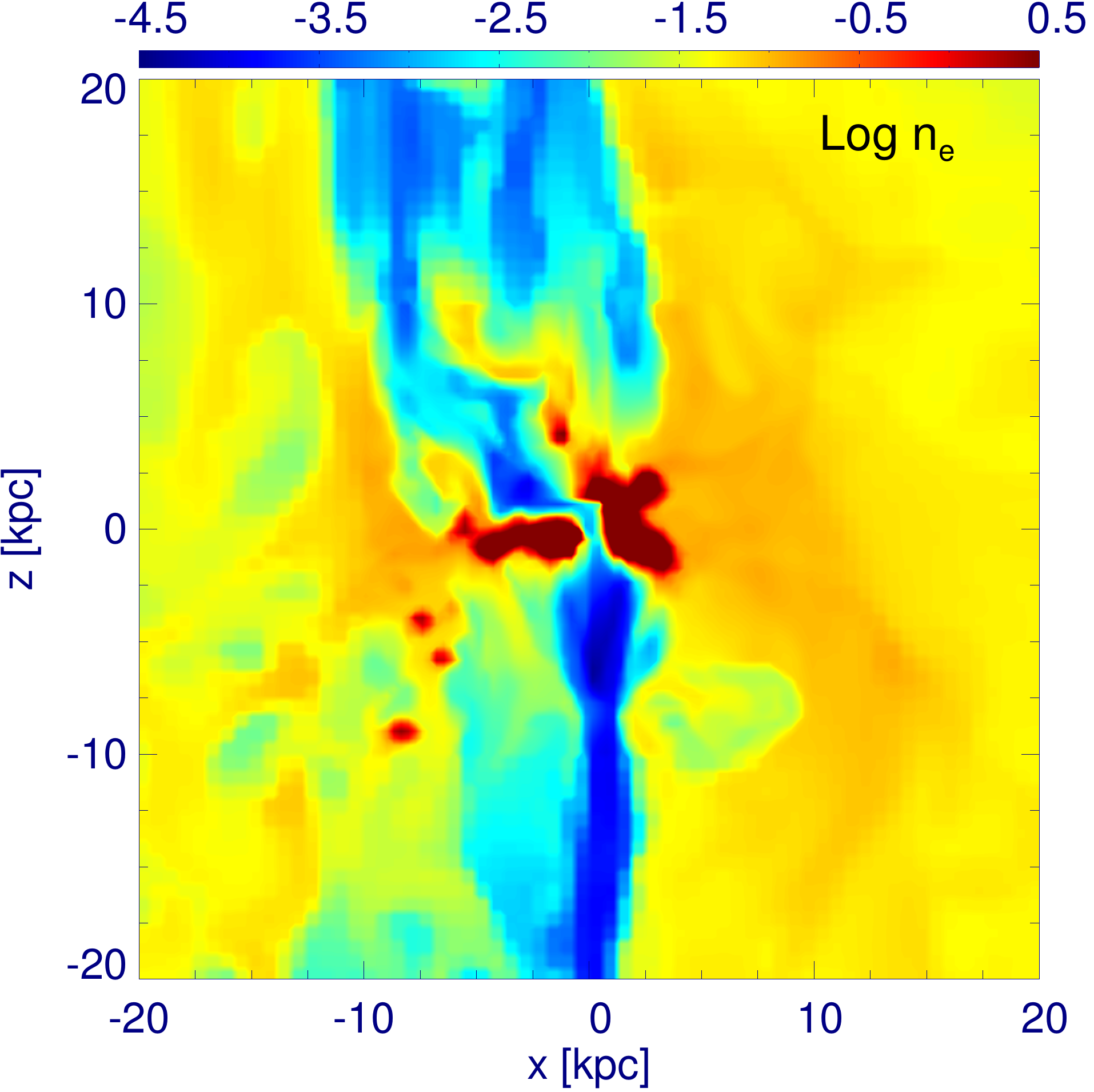}}
    \subfigure{\includegraphics[scale=0.31]{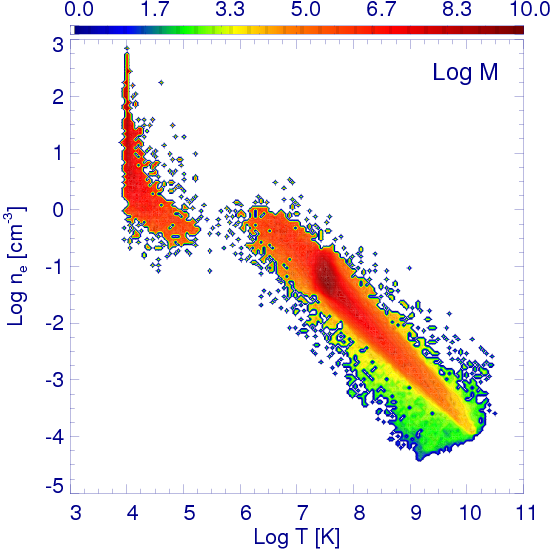}}
     \caption{Same as Figure \ref{fig:7_6em3_maps} but for AGN feedback efficiency $\epsilon=10^{-2}$ (r7-1em2).
     The time of the maps is 0.61 Gyr. 
     Notice that concentrated cooling, in the form of
     a rotating cold torus, is favored over extended cold gas
     ($t_{\rm cool}/t_{\rm ff}$ is usually $>10$). 
     \\
     \\
      \label{fig:7_1em2_maps}}
\end{figure*}

We now discuss a run for higher mechanical efficiency ({$\epsilon=10^{-2}$) compared to the one considered in the previous Section ({$\epsilon=6\times 10^{-3}$). 
We note that these values likely bracket the range of acceptable efficiency levels,
as pointed out by previous computations (G11a; S11) and observations (\citealt{meh08}).
We use the same 
set of diagnostic plots as before and we only discuss the differences between these models.
The results for $\epsilon=10^{-2}$ are shown in Figures \ref{fig:7_1em2_phases_jet} and \ref{fig:7_1em2_maps}, which are direct 
counterparts to Figures \ref{fig:7_6em3_phases_jet} and \ref{fig:7_6em3_maps}, respectively.

The increase in the mechanical efficiency of the jets leads to stronger 
outflows 
characterized by a slightly more peaked jet power distribution centered around $\sim6\times10^{44}$ erg s$^{-1}$. Because in the adopted feedback scheme $\dot{M}_{\rm jet}$ is proportional to the jet power, the outflows also reach slightly higher mass loading rates 
($\ga 0.5$ $\msun$ yr$^{-1}$). 
This means that some jet events 
carry larger amount of linear momentum and, thus, have more `piercing power'
with which they plow through the ICM. Consequently, the jets release more of their energy at relatively larger distances from the cluster center, allowing a bit more cold gas accumulation along the equatorial region. The projected surface brightness map clearly shows the formation of a torus-like structure that was less pronounced in the lower efficiency run.

The initial increase in the average cooling (and accretion) rate is strongly quenched at around 1 Gyr since the beginning of the simulation, which is earlier than in the previous run. 
After this short phase, the system gradually 
settles to a quasi equilibrium state, with cooling rates around $40-50$ $\msun$ yr$^{-1}$, twice the level seen in the lower efficiency case. Paradoxically, more powerful jets are less efficient in stopping the cooling flow, due to tunnel carving. The total injected energy is comparable in both runs ($\sim10^{62}$ erg), a sign that the self-regulation process is overall very similar, although with a slightly shorter 
active phase and a lower frequency of the outbursts (see jet power CDF).  The total cold gas mass ($2\times10^{11}$ $\msun$)
is slightly larger than in the preceding case, but still only $10\%$ of the pure cooling flow value, which is consistent with the observational constraints.

During the entire evolution of this simulation, the outflows tend to produce bigger bubbles or a deeper channel (see also the $n_{e}$ cut).
The effect of more powerful jets is also to increase the scatter in the phase diagram: 
the gas in hot phase occupies a wider range of values in the $n_{e}-T$ plane 
about the (imaginary) diagonal line representing average pressure.

The minimum TI-ratio falls below 10 only early in the evolution. 
Extended cold gas, in the form of blobs and filaments,
is mainly formed at that epoch (see the $n_{e}$ cut).
At later times, more powerful events raise $t_{\rm cool}/t_{\rm ff}$ well over 10, and the ICM becomes hotter. 
Even if the system settles later around a minimum TI-ratio of 13, extended cold filaments may be 
occasionally seen at moderate radii 
due to dredge up of the cold torus by  outflows. 
The uplift of the cold gas and the expansion of bubbles may further increase density, possibly triggering new thermal instabilities.
This picture is also consistent with the observational analysis presented in M11 (see their Fig. 12, right panel), suggesting that 
value 10 marks an approximate transition to the regime where the multiphase gas may appear.

\subsubsection[]{$\epsilon=6\times10^{-3}$, Wobbling}

To test the robustness of our results we included jet wobbling in the run with mechanical efficiency $\epsilon=6\times10^{-3}$.
Observational data is consistent with the possibility that the jets may precess or their orientation
may change. Since the data is very limited, we keep the model very simple and
change jet orientation whenever the accreted mass reaches a threshold of $10^7$ $\msun$. For 
typical mass accretion rate of order $\sim 1$ $\msun$ yr$^{-1}$, this corresponds to a 
reorientation timescale of $\sim 10$ Myr (e.g., \citealt{dun06}). 

In the first simulation with jet wobbling, we assumed that the probability of 
jet pointing in a given direction is given by a random uniform distribution over 4$\pi$ steradians
(the bipolar jets are always collinear).
Initially, the jet orientation changes at a slower rate, $\sim 100$ Myr, because the accretion rate is relatively low. Thus, in this early stage, the evolution is very similar to that 
for the non-wobbling case with efficiency of $\epsilon=6\times10^{-3}$. However, the energy is 
better spatially distributed, and thus the cooling rate is more quenched,
reaching values $\la 10$ $\msun$ yr$^{-1}$. 
As the accretion rate accelerates, and the jet orientation starts to change on timescales of 
10 Myr or less, the deposition of energy becomes nearly isotropic. 
A more uniform distribution of energy makes it more difficult 
to feed the black hole via the cold torus, which is now strongly disturbed.
Rather than being used to drill a channel through the ICM,
the AGN jet is no longer able to efficiently deliver the energy to the ICM beyond $r>15$ kpc. 
This enables the gas to cool rapidly at these radii, which 
leads to the cooling catastrophe in about 1 Gyr. Larger efficiencies could possibly stop the 
cooling, but the nearly isotropic centrally-concentrated energy distribution would destroy the cool core.

In order to remedy this situation, we considered models where the jet orientation is confined to 
two conical regions of half opening angle 75$^\circ$, 45$^\circ$, and 25$^\circ$. 
As before, the probability distribution of the jet orientations is randomly uniform within these regions. 
In the first two cases, the final result is similar to the case considered above, but the cooling flow is quenched for an additional time of $\la 1$ Gyr.

For half opening angle of 25$^\circ$, the jet energy distribution becomes narrower and more anisotropic.
Apart from larger bubbles and a more frequent jet channel
fragmentation, this computation resembles the previous best model r7-6em3: the cooling rate is
steady for several Gyr ($\sim10-30$ $\msun$ yr$^{-1}$), with total cold gas mass around $10^{11}$
$\msun$. Cold blobs are produced mainly in the first 1.5 Gyr when TI-ratio oscillates
around 10. At later times, the system settles to a hotter state characterized by the TI-ratio of $\sim$15-16. 
Jet events are also more variable than in the r7-6em3 model, changing rapidly
from $10^{43}$ to $10^{45}$ erg s$^{-1}$, because more inclined outbursts are able to partially
dismantle the central torus, temporarily inhibiting accretion.

In summary, randomly wobbling jets
with small and moderate opening angles produce similar results to unidirectional narrow outflows,
especially for stopping the cooling flow catastrophe and generating extended multiphase gas.
The other relevant feature is that the central accumulated cold torus can be frequently destroyed, 
depending on the wobbling parameters.\\

\subsection[]{AGN Feedback: $t_{\rm cool}/t_{\rm ff}=21$}
\subsubsection[]{$\epsilon=6\times10^{-3}$}

\begin{figure*} 
    \subfigure{\includegraphics[scale=0.453]{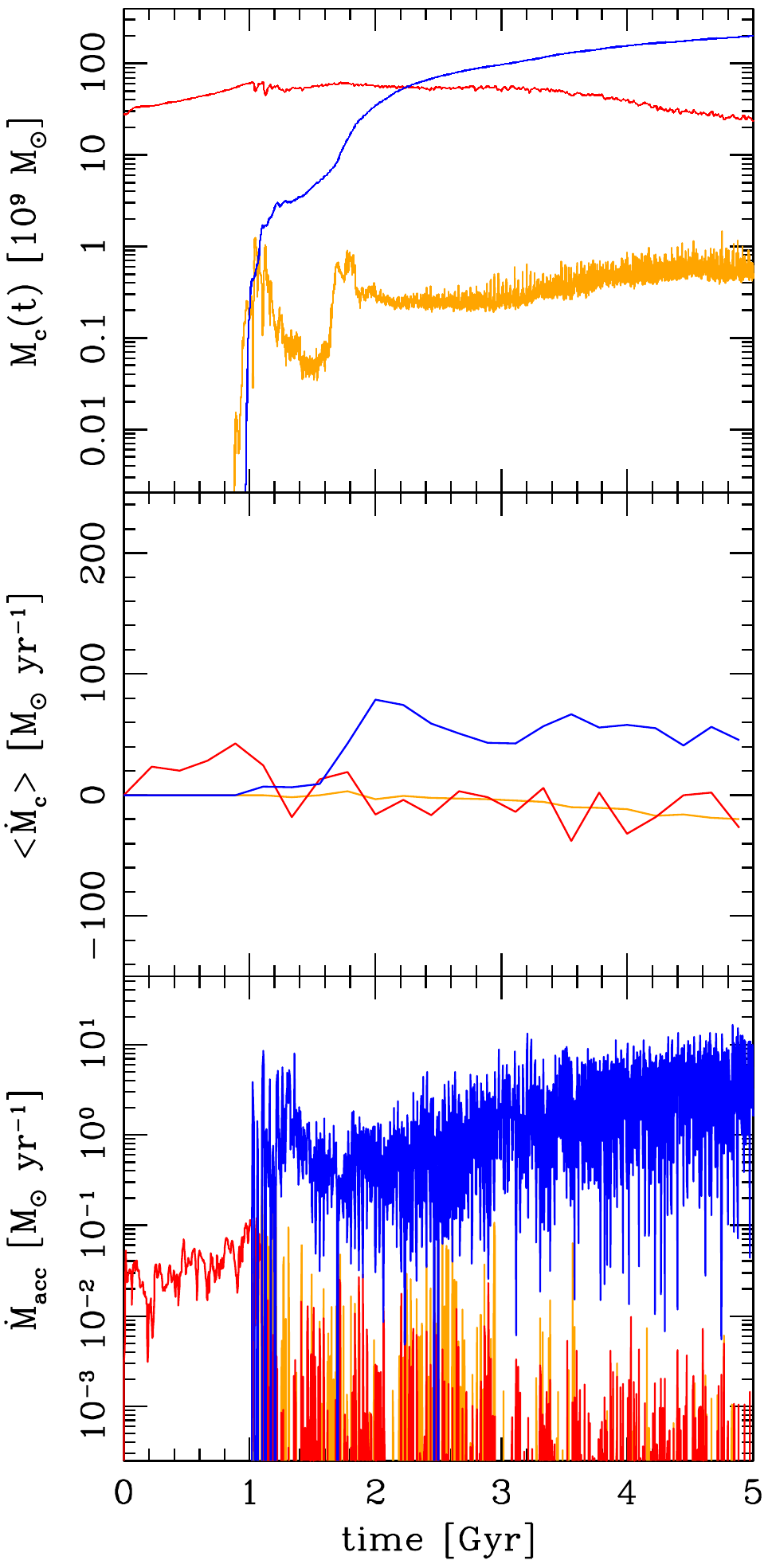}}
    \subfigure{\includegraphics[scale=0.453]{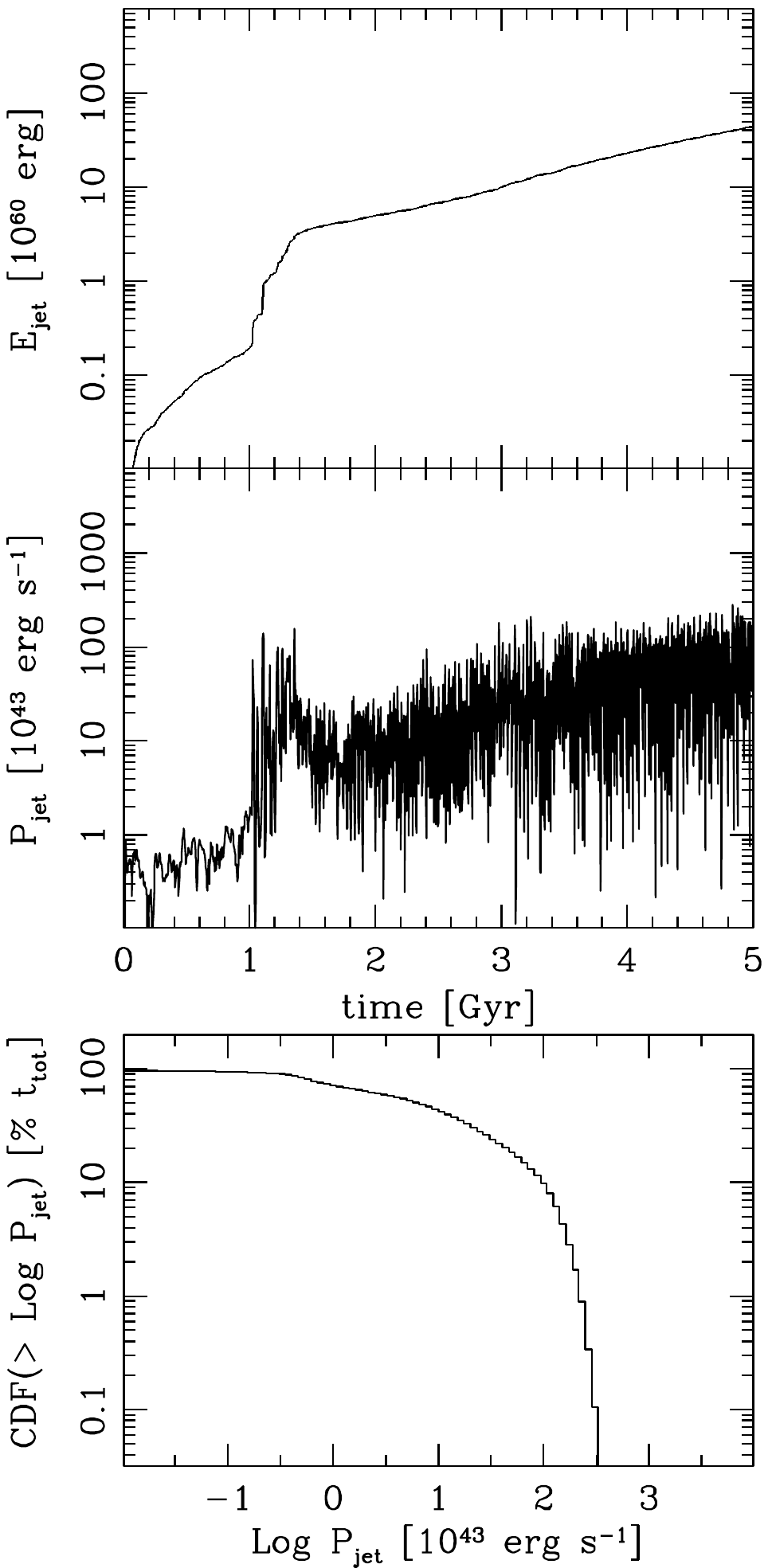}}
    \subfigure{\includegraphics[scale=0.453]{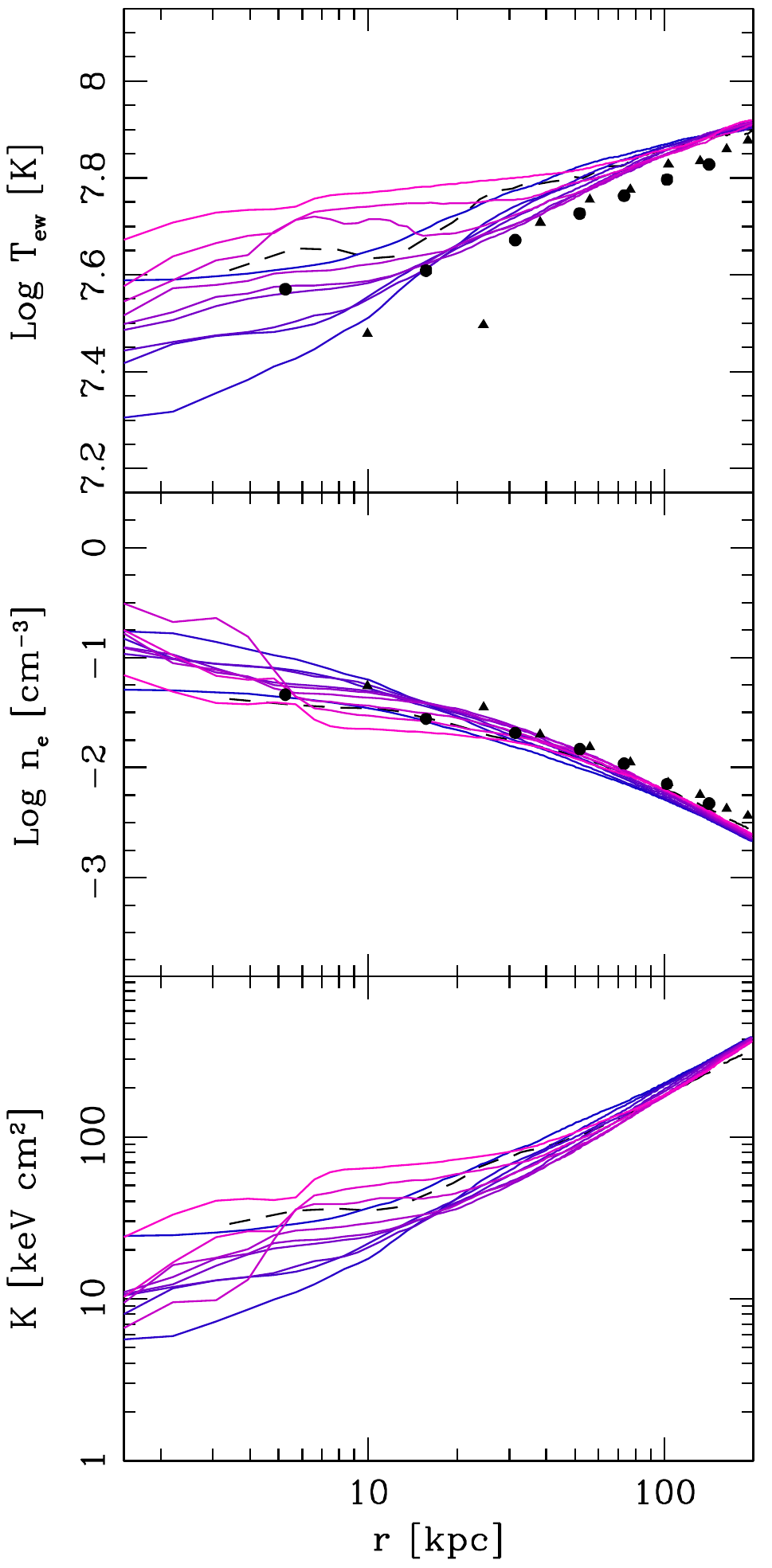}}
    \subfigure{\includegraphics[scale=0.453]{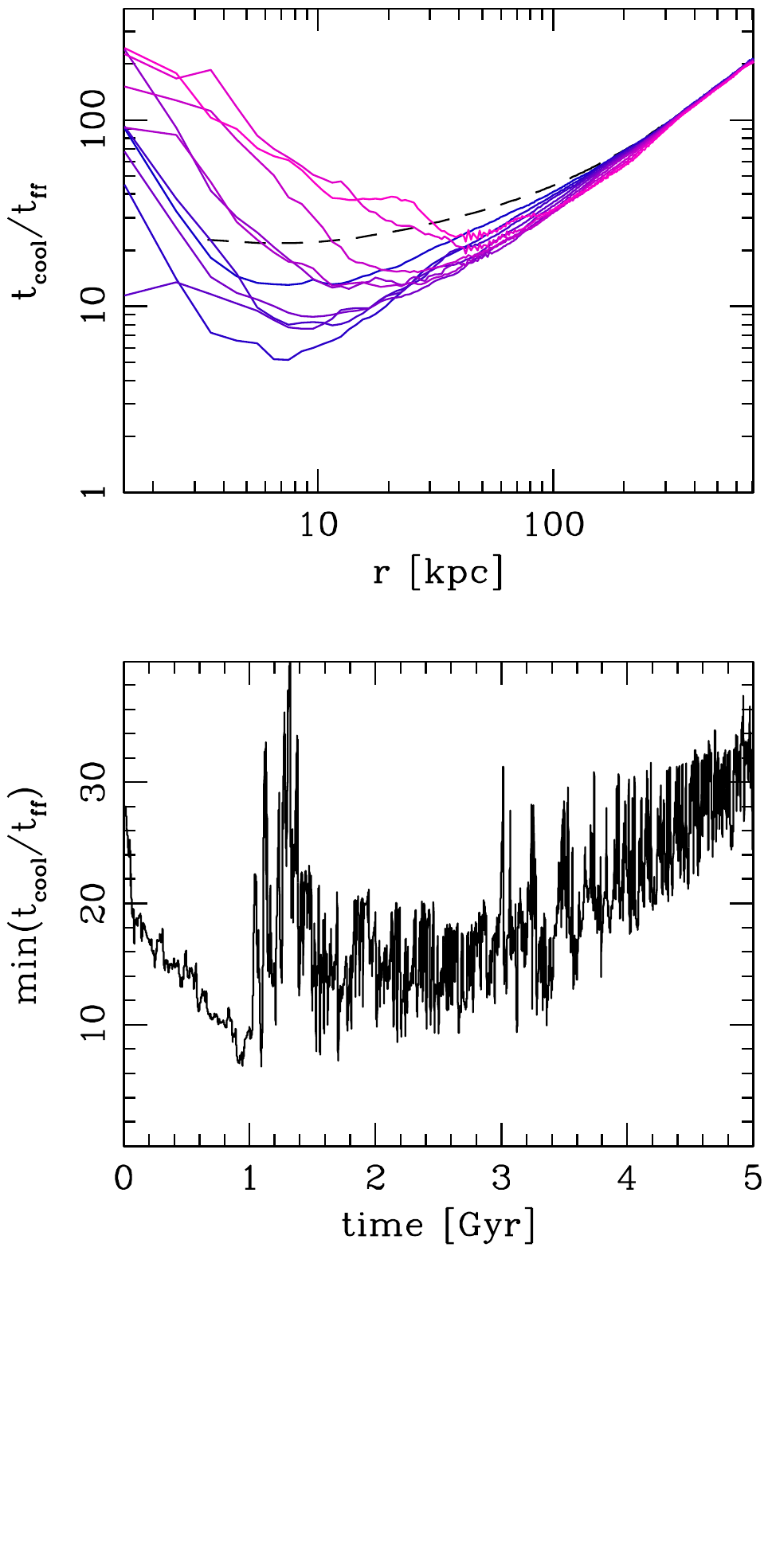}}      
     \caption{Same as Figure \ref{fig:7_6em3_phases_jet} but for initial minimum $t_{\rm cool}/t_{\rm ff}=21$ and
      AGN feedback $\epsilon=6\times10^{-3}$ (r21-6em3). 
     Starting with a higher TI-ratio produces an initial phase of slow cooling, 
     followed again by strong heating when the $t_{\rm cool}/t_{\rm ff}$ ratio approaches 10 (the multiphase gas threshold). 
     AGN outflows have a relatively stronger impact on the lighter ICM, secularly increasing
     the TI-ratio back above 20. This computation shows that the average thermal
     equilibrium is actually quasi stable, with phases of slight cooling or heating dominating throughout
     the entire evolution.
     \label{fig:21_6em3_phases_jet}
     }
\end{figure*} 
\begin{figure*}  
    \subfigure{\includegraphics[scale=0.31]{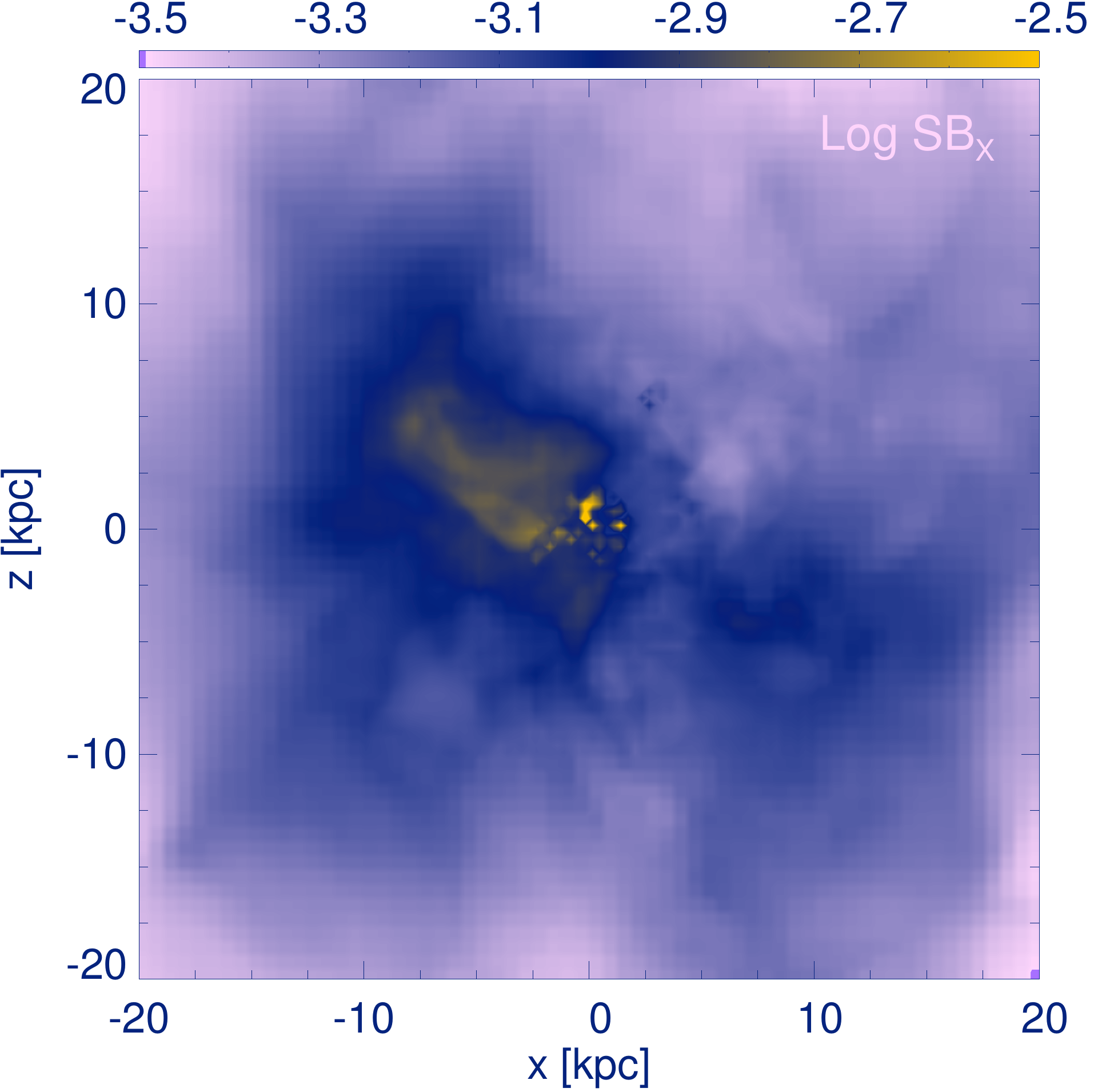}}
    \subfigure{\includegraphics[scale=0.31]{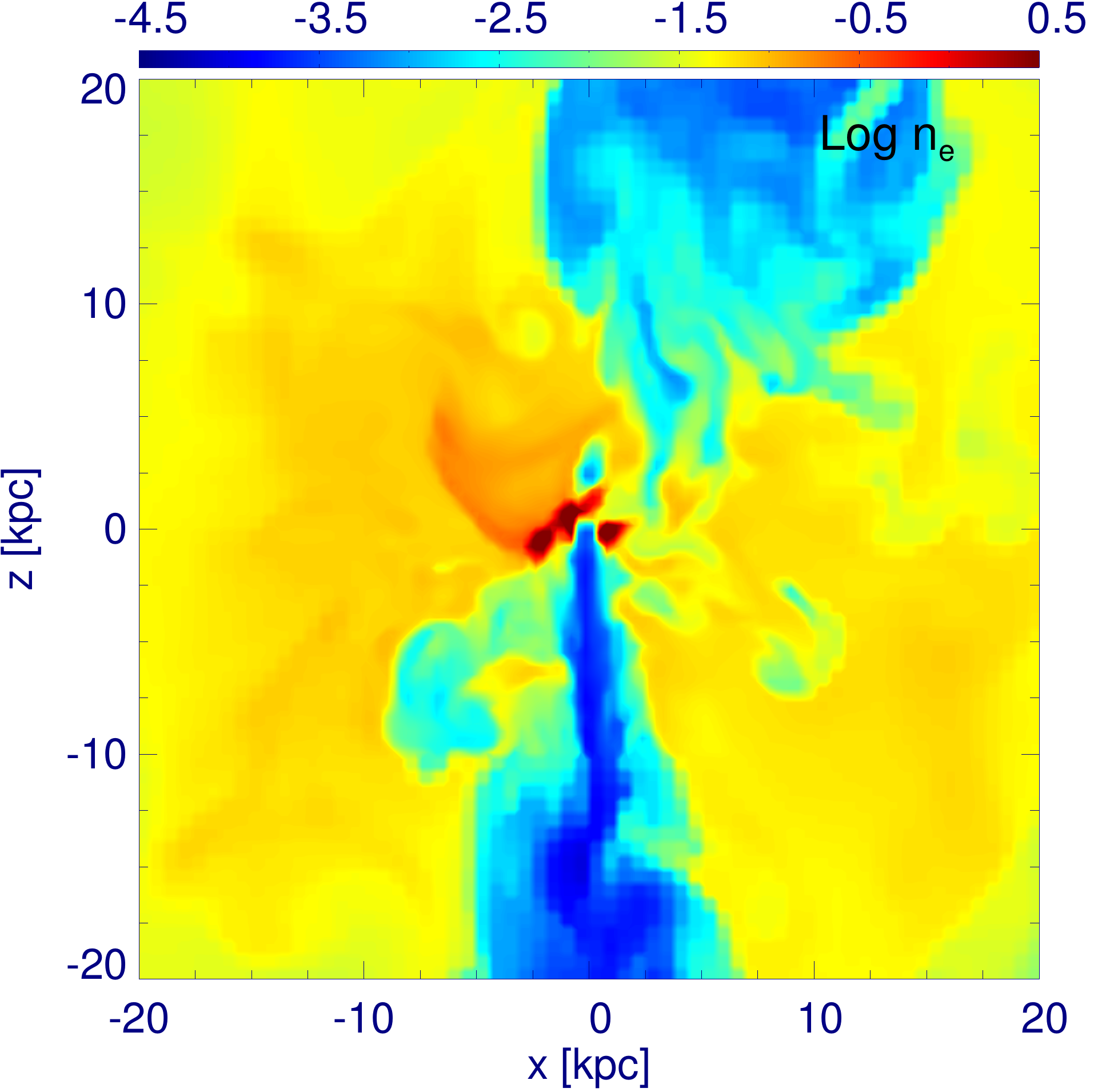}}
    \subfigure{\includegraphics[scale=0.31]{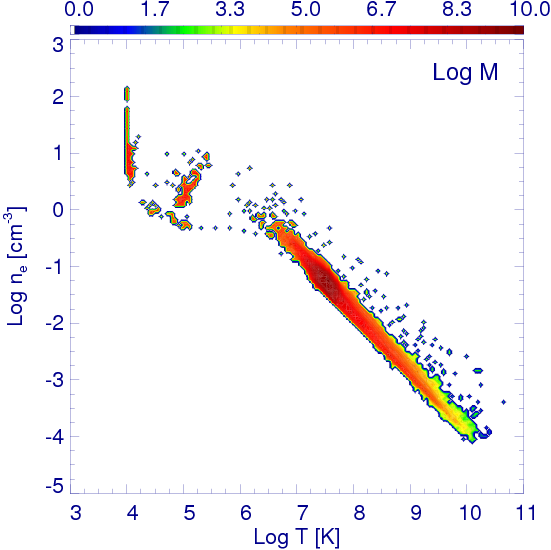}}
     \caption{Same as \ref{fig:7_6em3_maps} but for initial minimum $t_{\rm cool}/t_{\rm ff}=21$ and
     AGN feedback $\epsilon=6\times10^{-3}$ (r21-6em3). 
     The time of the maps is 1.27 Gyr. 
     Notice that concentrated cooling is favored through the whole evolution, while
     cold extended blobs condense out of the hot phase only in short windows (again whenever $t_{\rm cool}/t_{\rm ff}\la10$).
     \\
     \\
     \label{fig:21_6em3_maps}
      }
\end{figure*} 

\noindent
In order to study the effect of feedback in slower cooling systems, we generated new initial
hydrostatic conditions as follows. We decreased the central density preserving the initial profile slope at $r\ga 80$ kpc; simultaneously, we increased the central temperature to ensure that the ICM remains in hydrostatic equilibrium.
The new initial conditions correspond to the minimum TI-ratio of $\sim 21$.
The new initial profiles are plotted in Fig.~\ref{fig:21_6em3_phases_jet}, 
black dashed lines (third column). The entropy is elevated to 30 keV cm$^2$
at 10 kpc -- still within observational bounds (\citealt{cav09}) -- while the total gas mass 
is decreased only by 8\%. The results from the runs corresponding to these new initial conditions and AGN mechanical efficiency of $\epsilon=6\times10^{-3}$ are shown in Figures \ref{fig:21_6em3_phases_jet} and \ref{fig:21_6em3_maps}. These figures are analogous to those previously discussed. 

Overall, the behavior of the system is quite similar to what we observed before. The model is successful
in predicting a globally stable ICM core, low cooling rates ($\sim50$ $\msun$ yr$^{-1}$, 10\% of pure CF), 
preservation of the cool core (e.g., entropy, SB$_{\rm X}$), typical features of AGN jets (bubbles, shocks, turbulence), and the amount of multiphase medium consistent with observations. 
The main result is
that the local thermal instability is triggered despite the fact that the {\em initial} minimum TI-ratio exceeds $\sim 10$. It is only when the instantaneous value of this ratio falls below the critical threshold of $\sim 10$ that the multiphase medium suddenly appears in the simulation (e.g., after 1 Gyr). 
As before, this leads to increased feedback that brings the system
back to the high TI-ratio regime, which in this case is well above $\sim 20$ and 
higher than in the previously considered systems. 

Even if the typical jet power (few $10^{44}$ erg s$^{-1}$)
and total injected energy ($4\times10^{61}$ erg; single jet) are lower compared to r7-6em3, the
AGN outflows have a relatively stronger impact on the lighter ICM, secularly increasing
the TI-ratio back above 20. As a consequence, spatially concentrated cooling, in the form of a rotating
torus, is present during the entire evolution.
In Section 4 we make an in-depth comparison between the
models with different initial $t_{\rm cool}/t_{\rm ff}$ ratio. 

\subsubsection[]{$\epsilon=10^{-2}$}
The total cold gas mass in the model discussed above reaches $2\times10^{11}$ $\msun$, which is higher than r7-6em3.
Although this is observationally acceptable, some systems show smaller amounts of cold gas mass.
Therefore, we experimented with increasing the mechanical efficiency in order to decrease the 
amount of cold gas (figures not shown). However, as argued in Sec.~3.2.2, higher efficiencies lead to 
stronger and more mass-loaded jets that deposit their mechanical energy farther away from the center.
If in the initial outbursts the system appears to be more heated due to 
shocks, 
the situation reverses later, when the jet carves a deeper channel in the ICM. Thus, 
more accretion occurs along the equator and a strong cooling flow develops.
As seen before, this feature could be alleviated by a randomly wobbling outflow, with a less anisotropic energy deposition.

Overall, the simulations with higher minimum initial TI-ratio 21, can again produce extended multiphase gas whenever the instantaneous $t_{\rm cool}/t_{\rm ff}$ falls $\la 10$. In fact, in the early phase of the evolution, when cooling slightly prevails over heating, the TI-ratio slowly decreases, crossing the
threshold in about a Gyr. Later, stronger outbursts set the system back to a hotter state and higher TI-ratio, until the cycle restarts again (it may require several Gyr).
The main trend for the entire 5 Gyr is therefore concentrated cooling occurring in the core.

Contrary to the more massive models (TI-ratio 7), the right efficiencies in order to quench the cooling flow and preserve the cool core 
should not be more than $6\times10^{-3}$. 
A less massive system is in fact more susceptible to mechanical outbursts, and requires a more delicate feedback, in analogy to the environment of galaxy groups or ellipticals, as has been shown in previous works (G11b, S11).\\

\section{Discussion}

Galaxy clusters are extremely important astrophysical objects. Lying on the exponential tail of the halo mass distribution, they provide a sensitive probe of cosmology. They are also home to the most massive galaxies and black holes in the universe. Galaxy clusters are observed in all wavebands, from radio to X-ray and gamma rays. Owing to their deep potential wells, they are almost closed systems with the baryon fraction similar to the universal value. Moreover, most of the baryons in clusters are in the form of hot X-ray emitting plasma, the intracluster medium. 
Understanding the state of the ICM will not only make clusters more useful cosmological probes, but also shed light on galaxy formation and on the state of most of the baryons in the universe.

In this paper we try to understand the thermal state of the ICM core, closely tying our models with observations. From multiwavelength observations it is clear that the dense plasma with short cooling times in cluster cores is not cooling in a catastrophic fashion. What is the mechanism at the base of such self-preservation behavior?
Radio bubbles and X-ray cavities associated with cool-core clusters suggest that the
active galactic nucleus acts as a thermostat, fueled by the cooling gas.
As argued by G11a, the best realistic model able to maintain quasi thermal equilibrium for several Gyr, i.e., quenching the cooling flow without destroying the cool-core structure, is an asymmetrical input of mechanical energy, in the form of massive outflows. 
The cooling flow solution (Sec.~3.1) is very different from the observationally supported picture where the cluster core is sustained by AGN heating. Mass accretion and cooling rates in our best feedback simulations (Sec.~3.2.1 and 3.2.2) are suppressed by a factor $\gtrsim 10$ compared to a pure cooling flow, in line with observations (e.g., \citealt{pet01,pet03,tam03,pef06}). The X-ray surface brightness is also not strongly peaked in the center; the entropy and emission-weighted temperatures do not rapidly decline under the initial values, conserving the positive gradient. Using mechanical efficiencies in the range $5\times10^{-3}-10^{-2}$, the typical outflow powers are $\sim10^{44}-10^{45}$ erg s$^{-1}$, comparable to the X-ray luminosity.

The fundamental result, consistent with observations, is that extended cold filaments and blobs
condense out of the hot phase whenever $t_{\rm cool}/t_{\rm ff} \lesssim 10$ or, equivalently, when the central entropy becomes $\lesssim 30$ keV cm$^2$. 
The cold clumps fall then toward the center in about a free-fall time, 
forming a central cold rotating torus and in part boosting the black hole accretion rate.
The formation of cold filaments is a result of interplay between thermal instability and gravity. 
Previous idealized analyses by M11 and S11 enforced strict thermal equilibrium and found that multiphase gas is generated according to the same threshold\footnote{The criterion in spherical coordinates is less stringent compared to the Cartesian case ($\sim 1$) because instabilities are amplified by geometrical compression, as blobs sink toward the center.}. 

A realistic feedback
displays a variegated self-regulated evolution, producing complex observed phenomena, which
are not captured by simple idealized prescriptions. In the present work, we tested bipolar AGN jets, tied to the mass accretion rate at very small radii\footnote{Our simulations do not resolve the sphere of influence of the SMBH, so the actual mass accretion rate might be smaller. Thus, in reality the feedback efficiency may be even larger. More work is needed to satisfactorily study the fate of the accreting cold gas.}.
The asymmetrical input of energy produces strong elliptical shocks and regularly inflates hot
bubbles, which rise  
outwards. At the same time the inflow is not entirely stopped: along
the equatorial direction the gas, mainly cold, continues to fuel the black hole. 
Turbulence, entrainment and mixing dissipate the directional kinetic energy of the jets, further inhibiting the cooling flow near the center. When the jet ram pressure is substantial,
the cold gas is also dredged up, as observed in a few cases such as Perseus and Hydra (\citealt{hat06,git11}). The cycle repeats, when cooling starts to dominate again and the cold gas condensation ignites more powerful outbursts, setting the system in a slightly hotter state.

\begin{figure*}  
 \center
   \subfigure{\includegraphics*[scale=0.81]{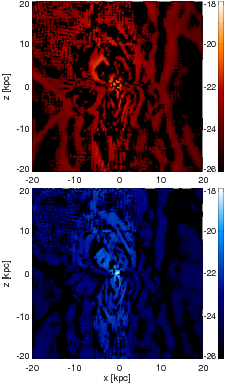}}
   \subfigure{\includegraphics*[scale=0.56]{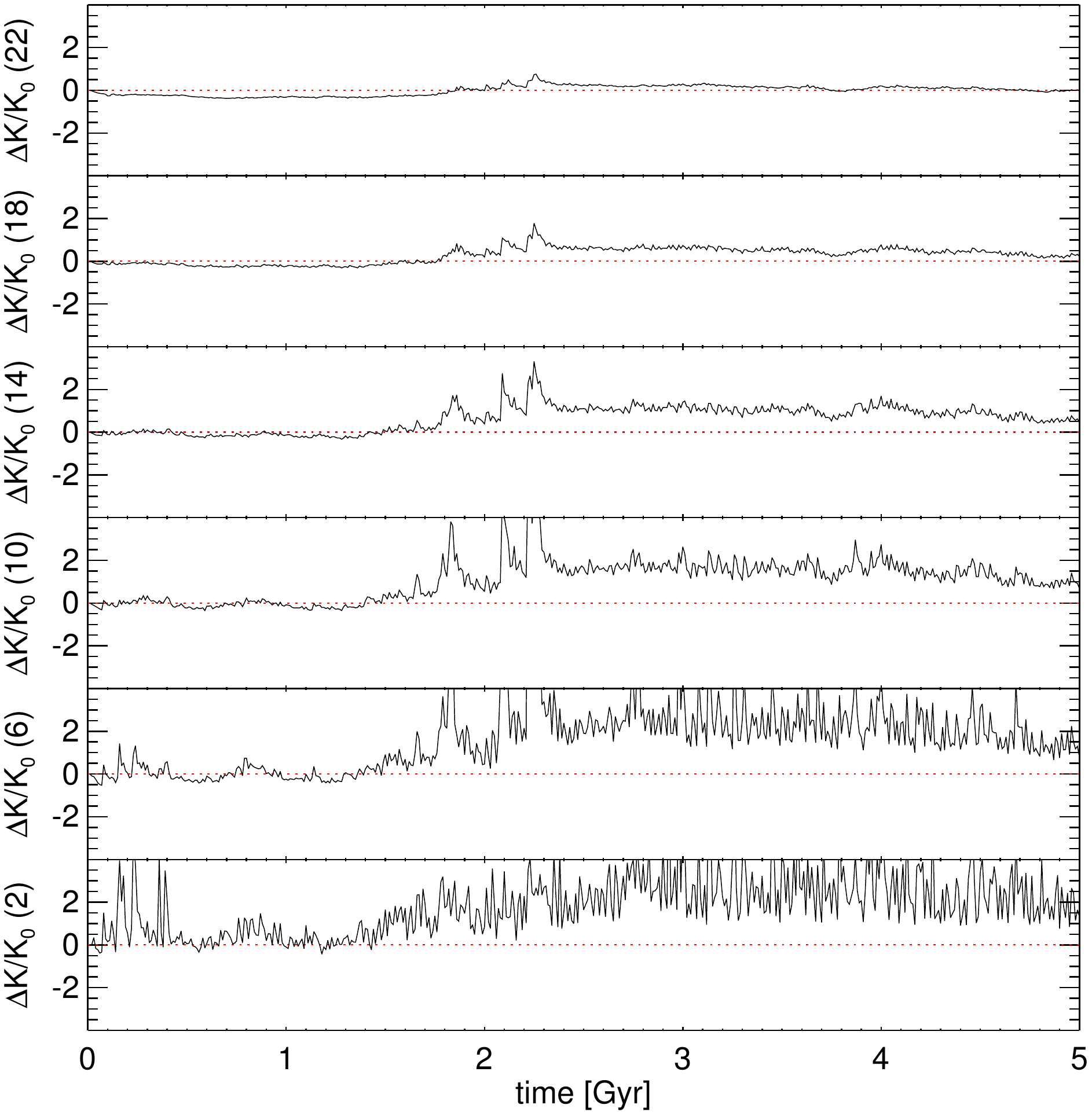}}
    \caption{Thermal equilibrium diagnostics (best model r7-6em3). Left:  2D midplane cut of the cluster core (20 kpc) showing the net internal energy increase (top) or decrement (bottom), averaged between 1 and 2 Gyr (erg s$^{-1}$ cm$^{-3}$; in logarithmic scale).
    Right: temporal variation of entropy of the whole ICM with respect to the initial value $K_0$.
    The six panels show ($K-K_0)/K_0$ as a function of time, 
    every 10 Myr, at different spherical shells
    of width 4 kpc centered at $r=22$, 18, 14, 10, 6 and 2 kpc (from top row). 
    \\
    \\
    \label{fig:TE}}
\end{figure*} 

Overall, thermal equilibrium is only roughly maintained, on large spatial and temporal scales. 
For example, the two maps in Figure \ref{fig:TE}, for best model r7-6em3, display the net thermal energy increase (red) or decrement (blue) per 
volume and time. The internal energy equation is given by
\begin{equation}\label{etherm}
\frac{\partial e}{\partial t} = -{\bf\nabla}\cdot\left(e\, {\bf v}\right) -P({\bf\nabla}\cdot{\bf v}) -\: n_{\rm{e}} n_{\rm{i}} \Lambda(T),
\end{equation}
where $e$ is the internal energy density (erg cm$^{-3}$). 
The maps indicate that `heating'
($\partial e/\partial t >0$) and `cooling' ($\partial e/\partial t<0$) alternate
on scales of a few kpc. 
Averages of $\partial e/\partial t$ in radial shells of width $\gtrsim4-5$ kpc 
suggest that a rough thermal equilibrium is maintained, 
although cooling appears slightly dominant in the central 10 kpc, while 
the jets tend to deposit their energy at larger radii. Equation (\ref{etherm}) does not, however, fully capture the irreversible heating associated with turbulent mixing, shocks and, 
wave damping due to numerical effects.
A better direct indicator of the balance between cooling ($\mathcal{L}$) and heating ($\mathcal{H}$) 
is the variation of the entropy $S$ ($\propto {\rm ln}\, K$, where $K\equiv k_{\rm b}T/n_{e}^{2/3}$), which is 
insensitive to adiabatic compressions or expansions:
\begin{equation}
\rho T \frac{dS}{dt} = \mathcal{H} - \mathcal{L},
\end{equation}
where $d/dt$ is the Lagrangian derivative.

The right panels in Figure 9 show the time variations in radial shells of the  `astrophysical' entropy
$K$ for the whole ICM, with respect to the initial value $K_{0}$. 
As expected the entropy increases on a secular timescale
due the continuous action of the AGN feedback. On smaller temporal scales, e.g., a few typical cooling times ($100-200$ Myr), 
the amplitude of the fluctuations in the six thick
spherical shells (centered at $r=22$, 18, 14, 10, 6 and 2 kpc; shown from the top to bottom row, respectively) can reach 2 or 3 times the mean value, especially in the nuclear region.
Although the fluctuations in $K$ can be substantial, its mean saturates at a stable level.
This suggests that the cluster core does not {\em dramatically} deviate from the state of (quasi) 
equilibrium between cooling and heating, at least
on scales of tens kpc and hundreds of Myr.
This was also suggested by the low cooling/accretion rates compared to a pure cooling flow.
The assumption of strict global thermal equilibrium made in the work of M11 and S11
is therefore very idealized.
It is thus particularly
remarkable that our 
criterion for the formation of the multiphase gas is in good agreement with their result.}

Unlike in S11, the central kpc region is not excised, allowing the cold gas (either nuclear cooling or fallen blobs) to form an angular momentum-supported torus as observed in the core of M87 (e.g., \citealt{for94}).  Such a disk can be partially dismantled when the jet direction is perturbed by turbulence or by random jet precession (Sec.~3.2.3). Nevertheless, the central dense disk is a common feature among 
elliptical galaxies (e.g., \citealt{mac97,mab03,ver06}). Centrally concentrated atomic gas is also seen in the survey of \citet{mcd10,mcd11a}. The central disk is certainly a nursery for new stars.
Since the star formation efficiency is quite uncertain (few per cent, up to 50\%, with an average value of $15\%$, according to \citealt{mcd11b}) we decided to not model star formation via a sink term. Thus, the central accumulation of cold gas could be further reduced.

The presence of multiphase gas is clearly seen in the $n_{\rm e}-T$ joint distributions. The cold dense phase ($10^4-10^5$ K), in the upper left corner, is present in large quantities and detached from the hot rarefied gas ($>$ few $10^7$ K), with a lack of gas at intermediate temperatures. A recent H$\alpha$ survey (\citealt{mcd10,mcd11a}) has shown that extended cold filaments are found in $\gtrsim 35\%$ of all clusters. It is striking that some of their maps (see their Fig.~4) resemble our computed
morphologies, with extended blobs and filaments up to $\sim20$ kpc: 
e.g., Abell 0496, Hydra A, Abell 1644, Sersic 159-03, 
and Abell 1795 (the cluster on which our model is based). Another 
case with substantial extended H$\alpha$ emission is the Perseus cluster (\citealt{sal06,fab08}). 
Moreover, the total cold gas mass we obtain, $\sim10^{11}$ $\msun$,
is consistent with the constraints of \citet{edg01}, whose
observations indicate clusters harboring up to $3\times10^{11}$
$\msun$ of molecular gas. 

The white noise perturbations in the initial conditions are wiped out over a cooling time. In fact, in the pure cooling flow computation, multiphase gas is only possible in such restricted interval, while the secular evolution is dominated by monolithic cooling. On the contrary, AGN feedback drives and sustains the
fluctuations in the long term evolution, although the original seeds 
initially help to distribute better the jet energy due to the interaction of jet with the inhomogeneous ICM.
The thermal state of the cluster core depends mainly on the feedback efficiency:
with smaller $\epsilon$ the frequency of cold filaments condensing out of the hot phase is larger (Figs. 3, 5, 7). Nevertheless, 
higher initial $t_{\rm cool}/t_{\rm ff}$ (and lower density) generates extended multiphase
gas according to the TI criterion, but thermal equilibrium is reached at later epochs. It is
interesting to note that AGN feedback is more effective in lighter systems, requiring therefore lower efficiencies ($\la6\times10^{-3}$), as has been shown for galaxy groups (G11b, S11).

\begin{figure} 
    \center
    \includegraphics*[scale=0.318]{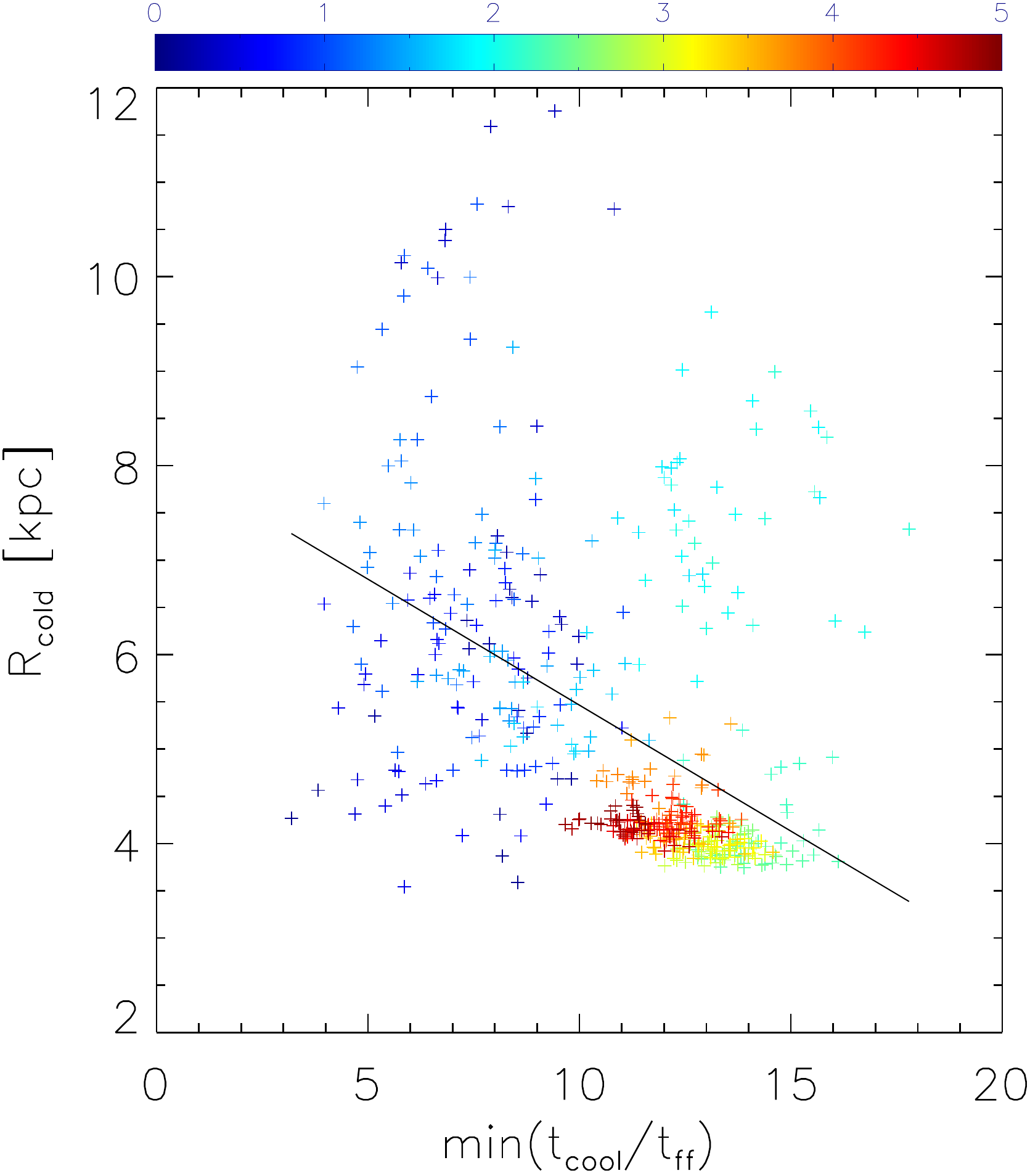}
    \includegraphics*[scale=0.318]{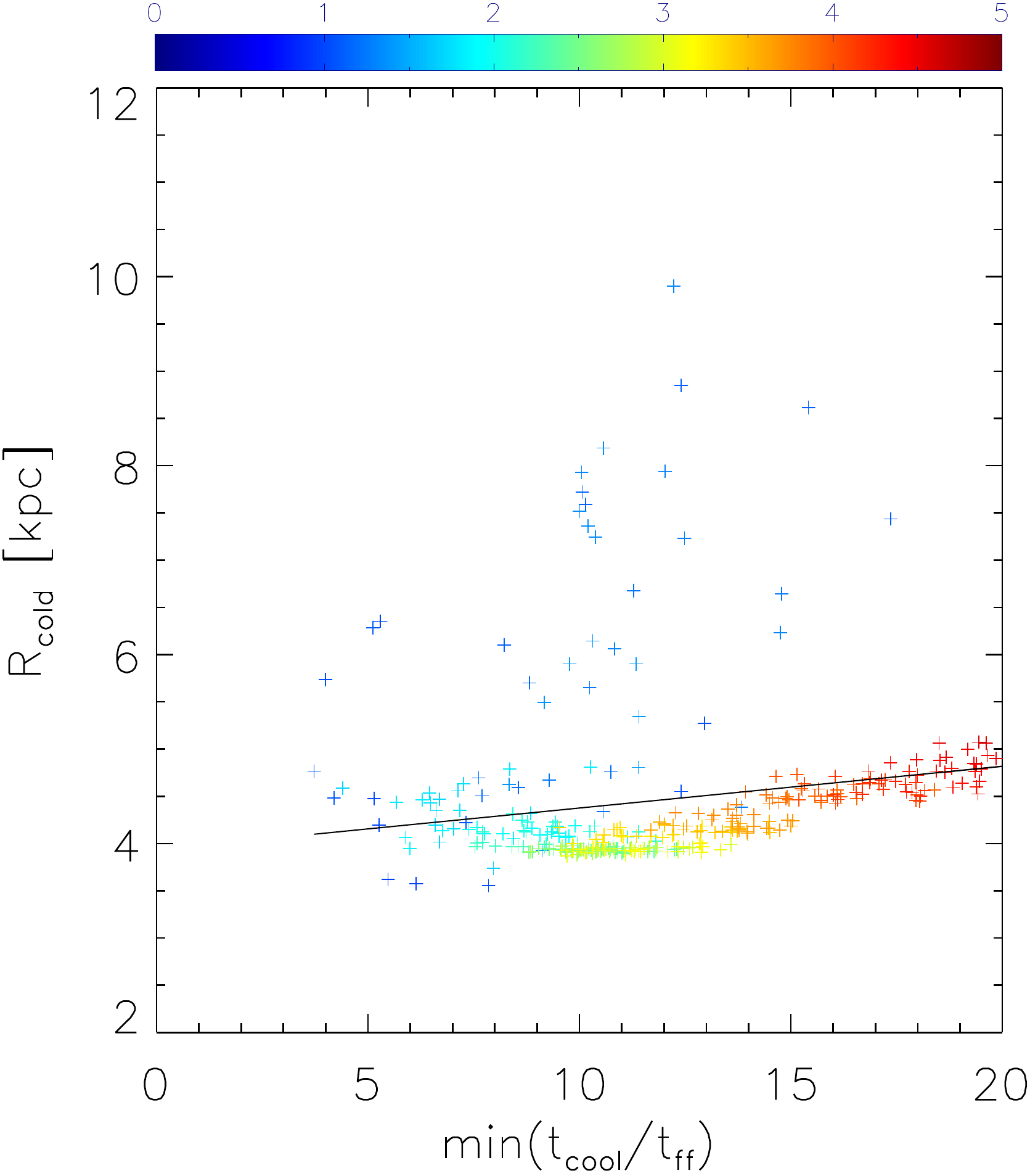}
     \caption{Extended cold radius as a function of minimum shell-averaged TI-ratio. This radius is 
     defined as the average distance from the center of cold cells
     that do no belong to the bulk of concentrated cooling ($r<3.5$ kpc). 
     The points are color coded according to time (Gyr; see color bar)
     Up: r7-6em3 simulation.
     Down: r21-6em3 simulation. Black line represents
     the linear regression fit to the scatter plot (points are drawn every 10 Myr). 
     Notice that these plots only show where the 
     cold gas instantaneously resides (on average), not where it originally formed. 
     \\
     \\
     \label{fig:Rcold_ratio}}
\end{figure} 
\begin{figure}    
   \center
    \includegraphics*[scale=0.318]{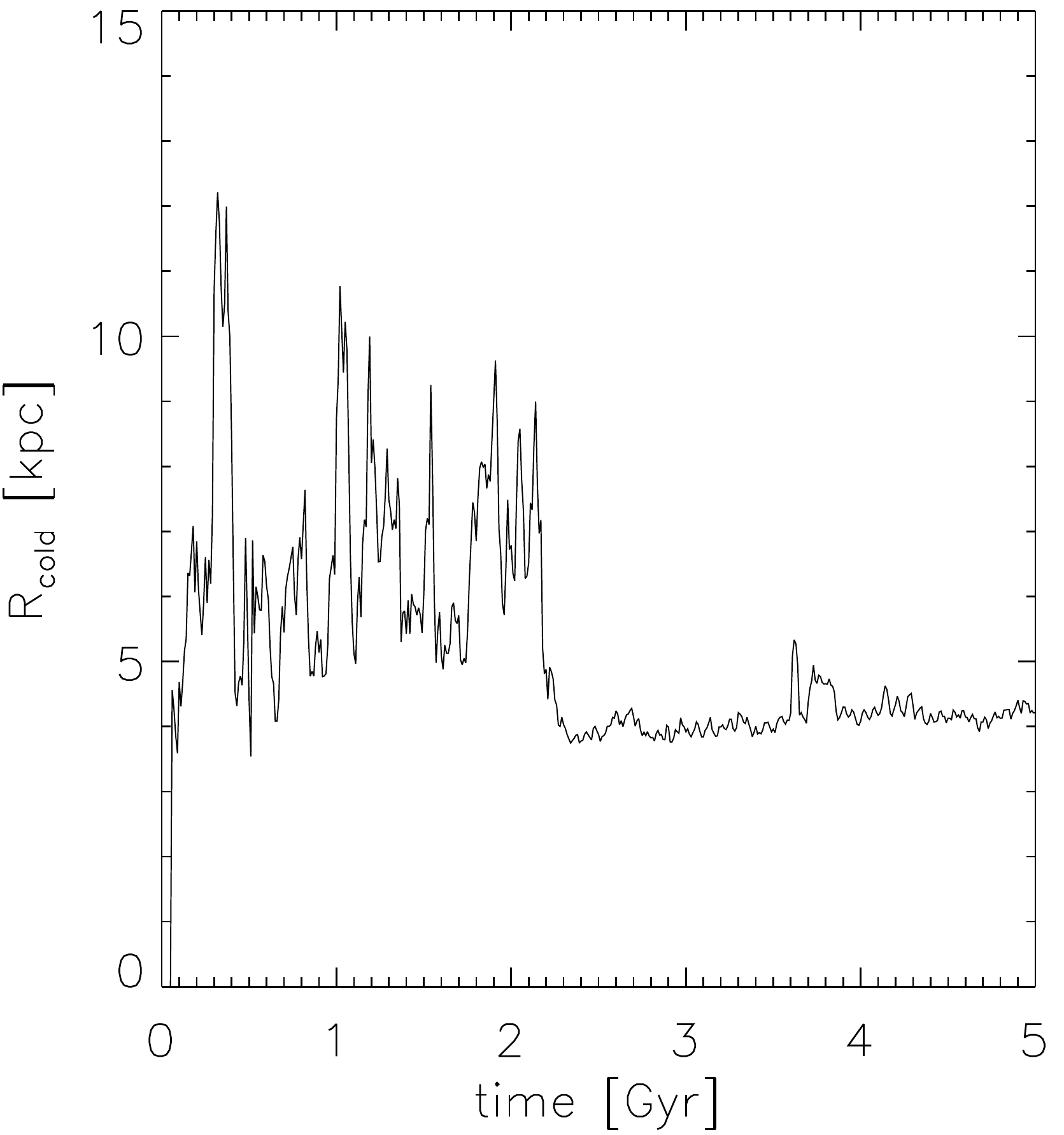}
    \includegraphics*[scale=0.318]{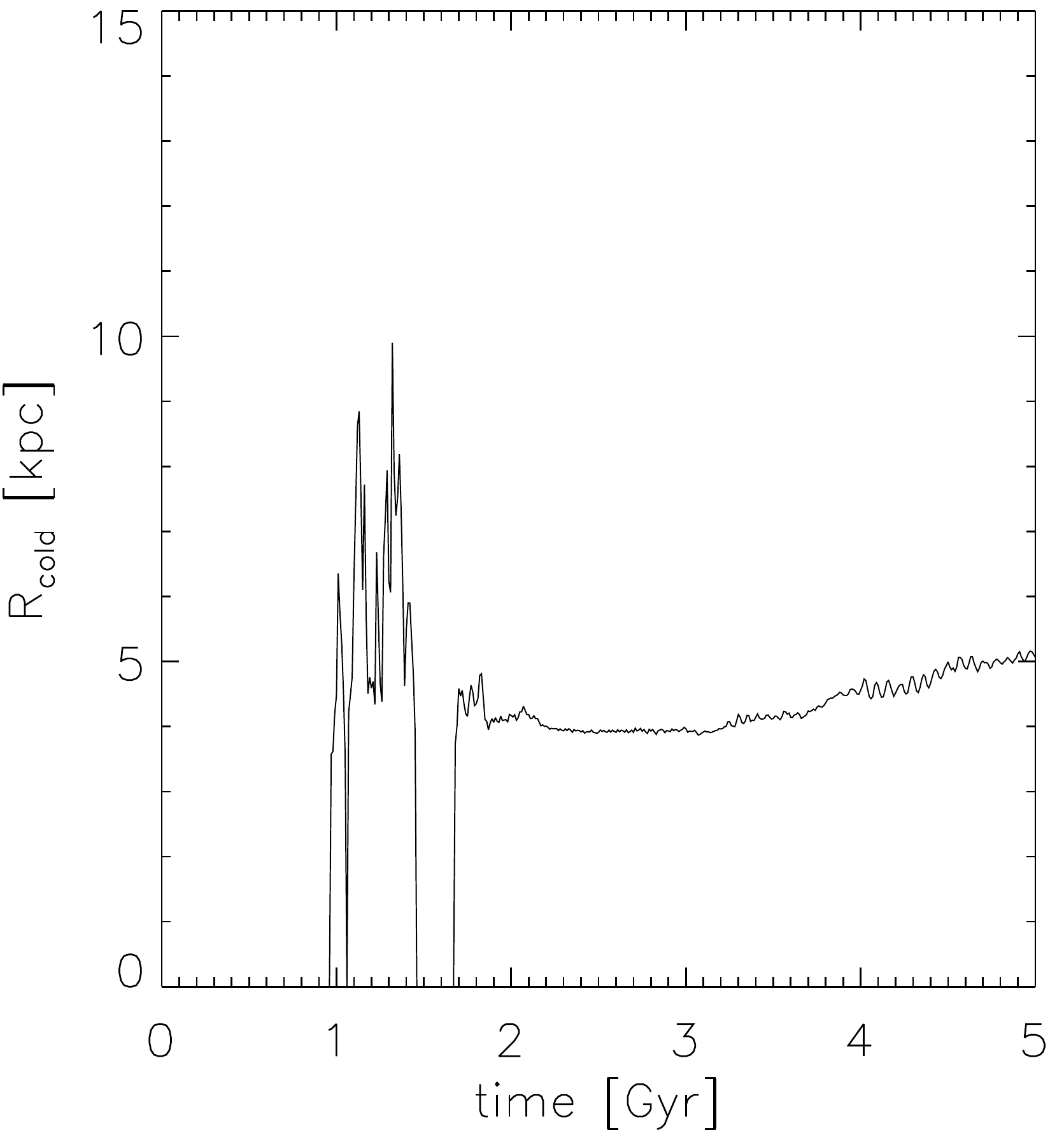}
     \caption{Extended cold radius as a function of time. Up: r7-6em3.
     Down: r21-6em3. Starting with lower TI-ratio leads to an evolution
     characterized by both extended ($t_{\rm cool}/t_{\rm ff}\la10$) and concentrated cold gas.
     On the contrary, extended multiphase gas is rarer in the opposite starting regime,
     although not absent.
     \\
     \\
     \label{fig:Rcold_time}}
\end{figure} 

To better quantify the presence of extended multiphase gas, Figure \ref{fig:Rcold_ratio} shows a scatter plot of the average distance of cold gas ($R_{\rm cold}$; excluding $r<$ 3.5 kpc) as a function of $t_{\rm cool}/t_{\rm ff}$. The points are color coded according to time: from blue (early) to red (late).
For the run with the initial $t_{\rm cool}/t_{\rm ff}=7$ (upper panel), the cold gas is much more spatially extended, especially in the first 2 Gyr. The run with less dense initial conditions (TI-ratio 21) shows a lower frequency of cold blobs at large distances because $t_{\rm cool}/t_{\rm ff} \gtrsim 10$ most of the time (compare Figs. 3 and 7). 
In both cases, the points clustered around $\sim 4$ kpc correspond to the rotating cold torus.
Since this plot shows the current average distance rather than the radius where the cold gas originates, 
extended filaments appear even when TI-ratio is elevated to $\sim 12-15$ due to AGN heating. 
This tendency is further enhanced if cold blobs are dredged up by powerful outflowing jets.
Moreover, as the TI-ratio is averaged in spherical shells, 
a few cold zones may contribute to $R_{\rm cold}$, without changing much the average $t_{\rm cool}/t_{\rm ff}$.
That is, some extended H$\alpha$ emission is consistent with $t_{\rm cool}/t_{\rm ff}$ slightly greater than 10.  

Figure \ref{fig:Rcold_time} displays $R_{\rm cold}$ gas as a function of time. With denser initial conditions (r7-6em3), $t_{\rm cool}/t_{\rm ff}$ is less than 10 for several short intervals within 2 Gyr and we expect extended cold gas\footnote{Note also that $\dot{M}_{\rm acc}$ is boosted exactly during this period (see Fig.~\ref{fig:7_6em3_phases_jet}).}.
Indeed, the average distance of the cold gas is significantly larger than that of the cold torus ($\approx 4$ kpc) during this period. Similarly, for the lower initial density run we see extended cold filaments at $1-1.5$ Gyr, again when $t_{\rm cool}/t_{\rm ff} \lesssim 10$;
later times are however dominated by the nuclear cold gas.
We conclude that the instantaneous value of $t_{\rm cool}/t_{\rm ff}$ is a robust indicator of extended cold gas formation, while nuclear cold gas can be present irrespective of the TI-ratio (e.g., bottom panel of Fig.~\ref{fig:Rcold_ratio}).

It is likely that other physical processes play an important role in shaping the morphology of condensing multiphase gas. \citet{sha10} and M11 showed that anisotropic thermal conduction leads to cold gas stretched along the local magnetic field direction. In order to be able to fully understand the exact topology of cold filaments, this physics should be included in the simulations (with an adequately higher resolution). 
However, the generation of cold filaments, whenever $t_{\rm cool}/t_{\rm ff} \lesssim 10$, is not likely 
to be affected by the presence of anisotropic thermal conduction.

The other physics that is missing in our simulations involves cosmological accretion and mergers, which drive additional turbulence, inducing more instabilities. Like AGN feedback, major mergers are associated with entropy generation and strong heating of cluster cores, raising the cooling time over the cluster age. Thus, mergers may be responsible for extreme non-cool-core clusters. However, they are also rare with timescales much longer than the cluster core cooling times (e.g., \citealt{fak10}). Therefore, AGN feedback, rather than mergers, is expected to be responsible for balancing cooling in most systems.

Our theoretical models are able to solve the cooling flow problem for several Gyr.
The key factor resides in adopting realistic mechanical AGN jets. The bulk of AGN heating is in fact a combination of shock dissipation and gradual thermalization of the mechanical energy
(see also \citealt{bru07}).
Random motions, either due to AGN, multiphase gas, and cosmological turbulence, strongly help the deposition of the feedback energy in the cool core, in contrast to idealized symmetric atmospheres (e.g., \citealt{vern06}). 

An important issue is whether our results depend sensitively on resolution. 
The morphology of the cold phase depends on the resolution. We 
note that finite resolution may prevent the disruption and mixing
of the small infalling clouds with the ambient ICM. Thus, limited resolution could in fact mimic 
some of the effects of the physics that we do not include (e.g., magnetic draping). 
We performed a test with two times lower resolution, but with the same resolution in the jet injection region (r7-6em3c),
and obtained very similar results. In particular, a steady global thermal equilibrium is reached in a few Gyr, with only 
a slightly larger final cold mass ($2\times10^{11}$ $\msun$). The evolution of the TI-ratio and condensed cold blobs appears also similar to the higher resolution model (r7-6em3), although the increased numerical dissipation tend to slow down
the growth of thermal instabilities at larger radii.
We note also that our simulations} have a resolution $\sim10$ times
higher compared to the work done in G11a. 
Although in G11a we adopted a cold mass dropout and a different jet injection method, AGN jets could also stop the cooling flow, in line with observations. Long term thermalization seems therefore not dramatically affected by resolution or the exact heating prescription. The fine details of the feedback can however change with higher resolution, e.g., jets tend to become more frequent (cf. G11a).\\

\section{Conclusions}

Our main conclusions can be summarized as follows:

\begin{itemize}
\item{Mechanical AGN feedback
self-consistently linked to BH accretion (with efficiency $5\times10^{-3}-10^{-2}$), 
can solve the cooling flow problem, 
quenching the cooling rate down to $\sim 10$ $\msun$ yr$^{-1}$ (1-10\% of the pure cooling flow value), 
preserving the cool core structure in agreement with observations (\citealt{pet01,pet03,ett02,tam03,pef06,mcn07}). 
Typical outflow powers are $10^{44}-10^{45}$ erg s$^{-1}$,
with velocities $\sim 10^4$ km s$^{-1}$ and mass loading rates
$\sim \msun$ yr$^{-1}$ (see \citealt{cre03,mor05,mor07,nes08,nes11,tom10}). 
Best models are associated with narrow jets with fixed inclinations
or fast random wobbling with half-opening angle $\la25^\circ$.\\} 

\item{Multiphase gas, with cold ($T\sim10^4$ K) and hot  ($T>10^7$ K) components, 
is a direct consequence and source of feedback. 
The cold gas comes in two phases: extended phase
(up to 10-20 kpc of distance form the center) 
in the form of blobs and filaments generated by thermal instabilities
(and occasionally due to dredge-up by jet ram pressure)
and a more spatially concentrated one 
in the form of central rotating torus due to nuclear cooling or accumulated fallen blobs.
The H$\alpha$ survey of \citet{mcd10,mcd11a} shows both  
kinds of H$\alpha$ emitting gas. Nuclear cold
reservoirs of gas, especially in the form of rotating disks are also a common feature among ellipticals 
(\citealt{for94,mac97,mab03,ver06}; a star
formation sink term and jet wobbling can reduce it considerably though).
Total cold gas masses are $\sim10^{11}$ $\msun$ after 5 Gyr, within the observational
constraints ($\la3\times10^{11}$ $\msun$, \citealt{edg01,sal03}).\\}

\item{The generation of extended cold gas is dictated by an important threshold:
whenever the instantaneous ratio of the cooling time to the free-fall time is $\lesssim10$ thermal
instabilities lead to cold gas condensing out of the hot ICM at large radii. 
On the contrary, when $t_{\rm cool}/t_{\rm ff}>10$,
slow runaway cooling occurs 
only 
within the central few kpc.
This threshold seems to be surprisingly robust, as predicted by simple analytical arguments (M11) and 
simulations with idealized feedback (S11). The criterion is also supported by recent observations 
(Fig. 12 in M11; \citealt{voi08,cav08}).\\}

\item{Multiphase ICM can exist in cool-core clusters irrespectively of their initial minimum 
$t_{\rm cool}/t_{\rm ff}$. With a realistic feedback the cluster core always finds a way to self-regulate toward 
a state of global {\it quasi} thermal equilibrium on large spatial and temporal scales. The system forgets about the initial state in a few cooling times and the eventual evolution depends primarily on the feedback efficiency. Lower $\epsilon$ results in frequent episodes of extended cold gas. 
Intense cooling episodes are followed by increased jet activity and the core returns to a higher entropy level. This phase is more prolonged and intense in less massive systems (e.g., initial $t_{\rm cool}/t_{\rm ff}=21$).
The reduced gas inflow leads to the decreasing jet power, allowing a slow spatially concentrated cooling to start again
($t_{\rm cool}/t_{\rm ff}>10$).\\}
\end{itemize}

Overall, the main thermal properties of cluster cores are determined by 
global quasi-thermal equilibrium, with an amazingly detailed complexity of cold filaments/torus, jets, bubbles and shocks. Additional processes, such as thermal conduction, magnetic fields and cosmological accretion, are required to make more precise comparisons with observations.
Nevertheless, we showed that multiphase gas due to thermal instabilities is an essential element of the feedback cycle: end product of cooling and fuel for the central black hole.
\\

\section*{Acknowledgments}
The software used in this work was in part developed by the DOE NNSA-ASC OASCR Flash Center at the University of Chicago.
We acknowledge the NASA awards SMD-10-1609, SMD-11-2209 (Pleiades),
the CINECA awards HP10BPTM62, HP10BOB5U6 (SP6), and
the University of Michigan for the availability 
of high performance computing resources and support. MG
would like to thank F. Brighenti for useful discussions and P. Temi, as the principal support scientist
at the NASA/Ames base. MR acknowledges the NSF grant 1008454.
Support for PS was partially provided by NASA through {\it Chandra} Postdoctoral Fellowship grant PF8-90054 awarded by the {\it Chandra} X-Ray Center, 
which is operated by the Smithsonian Astrophysical Observatory for NASA under contract NAS8-03060. We are also thankful for the hospitality of the Kavli Institute for Theoretical Physics (KITP) at UC Santa Barbara, where this work was initiated. 
\\


\label{lastpage}

\end{document}